\documentclass[aps,pra,twocolumn,groupedaddress]{revtex4}
\usepackage{graphicx}
\usepackage{multirow}
\usepackage{amsmath}
\usepackage{amssymb}
\usepackage[latin1]{inputenc}
\usepackage{array}
\usepackage{color}
\usepackage[caption=false]{subfig}

\begin{document}

\title{Faraday and Cotton-Mouton Effects of Helium at $\lambda = 1064$\,nm}
\author{A. Cad\`{e}ne$^1$}
\author{D. Sordes$^1$}
\author{P. Berceau$^{1}$}
\author{M. Fouch\'e$^{1}$}
\author{R. Battesti$^1$}
\author{C. Rizzo$^1$}
\email{carlo.rizzo@lncmi.cnrs.fr}

\affiliation{ $^{1}$Laboratoire National des Champs Magn\'etiques Intenses, (UPR 3228, CNRS-UPS-UJF-INSA),
31400 Toulouse, France.}

\date{\today}

\begin{abstract}
We present measurements of the Faraday and the Cotton-Mouton
effects of helium gas at $\lambda~=~1064$\,nm. Our apparatus is
based on an up-to-date resonant optical cavity coupled to
longitudinal and transverse magnetic fields. This cavity increases
the signal to be measured by more than a factor of 270\,000
compared to the one acquired after a single path of light in the
magnetic field region. We have reached a precision of a few
percent both for Faraday effect and Cotton-Mouton effect. Our
measurements give for the first time the experimental value of the
Faraday effect at $\lambda$=\,1064\,nm. This value is compatible with the theoretical prediction. Concerning Cotton-Mouton effect, our measurement is the second reported experimental value at this wavelength, and the first to agree at better than 1$\sigma$ with theoretical predictions.

\end{abstract}


\maketitle

\section{Introduction}

In 1845 Faraday discovered that a magnetic field affects the
propagation of light in a medium \cite{Faraday1846}. In
particular, he observed that a magnetic field parallel to the
light wave vector {\bf k} induces a polarization rotation of a
linearly polarized light. This effect is known nowadays as the
Faraday effect. With such experiments, Faraday was looking for the
proof that light and magnetic field have a common origin. These
revolutionary findings have been one of the most important steps
towards Maxwell's theory of electromagnetism.

At the very beginning of the 20th century, Kerr \cite{Kerr} and
Majorana \cite{Majorana} discovered that a linearly polarized
light, propagating in a medium in the presence of a magnetic
field, also acquires an ellipticity when the field is
perpendicular to {\bf k}. In the following years, this phenomenon
has been studied in details by A. Cotton and H. Mouton
\cite{CottonMouton} and it is known nowadays as the Cotton-Mouton
effect.

Faraday and Cotton-Mouton effects are both due to the fact that
the magnetic field creates an anisotropy in the medium which then
becomes birefringent. The term birefringent indicates that
different states of polarization do not have the same propagation
velocity. The Faraday effect corresponds to a magnetic circular
birefringence, i.e. the index of refraction $n_-$ for left
circularly polarized light is different from the index of
refraction $n_+$ for right circularly polarized light. The
difference  $\Delta n_\mathrm{F} = n_- - n_+$ is proportional to
the longitudinal magnetic field $B_\|$:
\begin{equation}
\Delta n_\mathrm{F} = k_\mathrm{F}B_\|,\label{Eq:kF}
\end{equation}
where $k_\mathrm{F}$ is the circular magnetic birefringence per
Tesla. On the other hand, the Cotton-Mouton effect corresponds to
a magnetic linear birefringence, $i.e.$ the index of refraction
$n_\parallel$ for light polarized parallel to the magnetic field
is different from the index of refraction $n_\perp$ for light
polarized perpendicular to the magnetic field. The difference
$\Delta n_\mathrm{CM} = n_\parallel - n_\perp$ is proportional to
the square of the transverse magnetic field $B_\bot^2$:
\begin{equation}
\Delta n_\mathrm{CM} = k_\mathrm{CM}B_\bot^2,
\end{equation}
where $k_\mathrm{CM}$ is the linear magnetic birefringence per
Tesla squared.

Such magnetic birefringences are usually very small ($\Delta
n_\mathrm{F}, \Delta n_\mathrm{CM} \ll 1$) for magnetic fields
available in laboratories, especially in the case of dilute
matter. Magnetic birefringence measurements are therefore an
experimental challenge. The value of the birefringence depends on
the microscopic matter response properties like
(hyper)susceptibilities. In the case of dilute matter, these
responses can be calculated {\it ab initio} using the
computational methods developed in the framework of quantum
chemistry \cite{Rizzo2005}. Experimental measurements are then a
fundamental test of our knowledge of the interaction of
electromagnetic fields and matter.

Among all known gases, helium presents the smallest Faraday and
Cotton-Mouton effects. {\it Ab initio} calculations of
the helium Faraday effect at $\lambda = 1064$\,nm, with $\lambda$
the light wavelength, have been published only
recently\,\cite{Ekstrom2005}. From the experimental point of view,
even if Faraday effect measurements in helium dates back to the
50s \cite{Ingersoll1956}, no measurement has yet been reported at
$\lambda = 1064$\,nm. Helium Cotton-Mouton effect has been first
measured at $\lambda = 514.5$\,nm in 1991 \cite{Cameron1991}. At
the same time, the first numerical calculation at a different
wavelength in the coupled Hartree-Fock approximation was published
\cite{Jamieson}. Actually, these two first values were not in
agreement. While some other theoretical calculations exist in
literature \cite{Rizzo1997}, only three more experimental values
have been published since 1991
\cite{Cameron1991,Muroo2003,Bregant2009}, with only one at
$\lambda = 1064$\,nm \cite{Bregant2009}.

{\it Ab initio} calculations of both Faraday and Cotton-Mouton
effect of helium are benchmark tests for computational methods. In
practice they can be considered as error free, especially when
compared with the error bars associated with the experimental
values. Experimental measurement precision has therefore to be as
good as possible to be able to test the different
computational methods.

Experimentally, one generally measures the Faraday effect by
measuring the polarization rotation angle $\theta_\mathrm{F}$,
related to the circular birefringence by the formula:
\begin{eqnarray} \label{Eq:ThetaF_intro}
\theta_\mathrm{F} = \pi \frac{L_B}{\lambda} \Delta n_\mathrm{F},
\end{eqnarray}
where $L_B$ is the length of the magnetic field region. The
Cotton-Mouton effect is measured through the induced ellipticity
related to the linear birefringence by the formula:
\begin{eqnarray} \label{Eq:Psi_intro}
\psi = \pi \frac{L_B}{\lambda} \Delta n_\mathrm{CM} \sin2\theta_\mathrm{P},
\end{eqnarray}
where $\theta_\mathrm{P}$ is the angle between light polarization and the magnetic field. Experiments are difficult because one needs a high magnetic field coupled to optics designed to detect very small variations of light velocity. One also needs a $L_B$ as large as possible. To this end, optical cavities are used to trap light in the magnetic field region and therefore increase the ellipticity to be measured (see e.g. Ref.\,\cite{Cameron1991}).

In this paper, we present measurements of the Faraday and the
Cotton-Mouton effects of helium gas at $\lambda = 1064$\,nm. Our
apparatus is based on an up-to-date resonant optical cavity coupled
to longitudinal and transverse magnetic fields. This cavity
increases the signal to be measured by more than a factor of
270\,000 compared to the one acquired after a single path of
light in the magnetic field region. This allows us to reach a
measurement precision of a few percent both for Faraday effect
and Cotton-Mouton effect. Our results are finally compared to the
theoretical predictions and they agree to within better than
1$\sigma$.

\section{Experimental setup and signal analysis}

\subsection{Apparatus}\label{Par:apparatus}

\begin{figure}[h]
\begin{center}
\includegraphics[width=8cm]{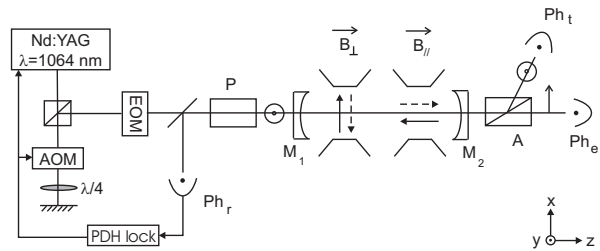}
\caption{\label{Fig:ExpSetup} Experimental setup. EOM = electro-optic modulator; AOM = acousto-optic modulator; PDH = Pound-Drever-Hall; Ph = photodiode; P = polarizer; A = analyzer. See text for more details.}
\end{center}
\end{figure}

Our apparatus is described in details in Refs.
\cite{Berceau2012,Battesti2008}. Briefly, as shown in
Fig.\,\ref{Fig:ExpSetup}, 30\,mW linearly polarized light provided
by a Nd:YAG laser ($\lambda = 1064$\,nm) is injected into a high
finesse Fabry-Pérot cavity consisting of the mirrors M$_1$ and
M$_2$. The laser frequency is locked onto the cavity using the
Pound-Drever-Hall method \cite{Drever1983}. To this end, the laser
passes through an electro-optic modulator (EOM) creating sidebands
at 10 MHz. The beam reflected by the cavity is detected by the
photodiode Ph$_\mathrm{r}$. This signal is used to adjust the
laser frequency with a bandwidth of 80\,kHz thanks to an
acousto-optic modulator (AOM) and with a bandwidth of a few\,kHz
thanks to the piezoelectric element of the laser. A slow control
with a bandwidth of a few\,mHz is also applied thanks to the
Peltier element of the laser.

Before entering the optical cavity, the light is linearly
polarized by the polarizer P. The light transmitted by the cavity
is then analyzed with the analyzer A crossed at maximum
extinction. Both polarizations are extracted: parallel and
perpendicular to P. The extraordinary beam (power
$I_{\mathrm{e}}$), corresponding to the light polarization
perpendicular to P, is collected by the low noise photodiode
Ph$_\mathrm{e}$, while the ordinary beam (power $I_{\mathrm{t}}$),
corresponding to the light polarization parallel to P, is detected
by Ph$_\mathrm{t}$. All the optical components from the polarizer
P to the analyzer A are placed in a high vacuum chamber which can
be filled with high purity gases. During this work,
magneto-optical measurements have been done using a bottle of
helium gas with a global purity higher than 99.9999\,$\%$. This
bottle is connected to the chamber through a leak valve allowing
to inject less than $10^{-3}$\,atm of gas.

Magnets providing a field perpendicular to the light wave vector
$\textbf{k}$ and a field parallel to $\textbf{k}$ surround the
vacuum pipe. The transverse magnetic field ($\textbf{B}_\perp\perp
\textbf{k}$) used for Cotton-Mouton effect measurements is created
thanks to pulsed coils described in Refs. \cite{Battesti2008,
Batut2008} and briefly detailed in section
\ref{Par:transverse_field}. For the Faraday effect measurements, a
modulated longitudinal magnetic field
($\textbf{B}_\parallel\parallel \textbf{k}$) is applied thanks to
a solenoid. More details are given in section
\ref{Par:longitudinal_field}.

\subsection{Fabry-P\'erot cavity}
A key element of the experiment is the Fabry-Pérot cavity. Its aim
is to accumulate the effect of the magnetic field by trapping the
light between two ultra high reflectivity mirrors M$_1$ and M$_2$.
The length of the cavity is $L_{\mathrm{c}}= (2.2713 \pm
0.0006)$\,m. This corresponds to a cavity free spectral range of
$\Delta^{\mathrm{FSR}}=c/2nL_\mathrm{c} = (65.996 \pm
0.017)$\,MHz, with $c$ the speed of light in vacuum and $n$ the
index of refraction of the medium in which the cavity is immersed.
This index of refraction will be considered equal to one. All
theses parameters and their uncertainties were measured
previously. Details concerning the measurement are given in
Ref.\,\cite{Berceau2012}. Using the Jones matrix formalism, we can
calculate the total acquired ellipticity due to the Cotton-Mouton effect $\Psi(t)$. It is linked to the ellipticity without any cavity
$\psi(t)$ by:
\begin{eqnarray}\label{Eq:Psi_psi}
\Psi(t) = \frac{2F}{\pi}\psi(t),\label{Eq:Psi_psi}
\end{eqnarray}
Likewise, the total rotation angle $\Theta_{\mathrm{F}}(t)$ due to the Faraday effect is:
\begin{eqnarray}
\Theta_{\mathrm{F}}(t) =
\frac{2F}{\pi}\theta_{\mathrm{F}}(t).\label{Eq:Theta_theta}
\end{eqnarray}
where $F$ is the finesse of the cavity and $\theta_{\mathrm{F}}(t)$ the rotation angle without any cavity.

\subsubsection{Cavity birefringence} \label{Par:bir_cavite}

The cavity induces a total static ellipticity $\Gamma$. This is
due to the mirrors intrinsic phase retardation \cite{Bielsa2009}.
Each mirror can be regarded as a wave plate and combination of
both wave plates gives a single wave plate. The total phase
retardation $\delta_{\mathrm{eq}}$ and the axis orientation of the
wave plate equivalent to the cavity depend on the phase
retardation of each mirror and on their relative orientation
\cite{Brandi1997,Jacob1995}. Thus the value of $\Gamma$ can be
adjusted by rotating the mirrors M$_1$ and M$_2$ around the
$z$-axis corresponding to the axis of light propagation.

We first set $\Gamma = 0$. To this end, we align the axis of the
equivalent wave plate on the incident polarization. This is done
by rotating the mirrors while the laser frequency is locked onto
the cavity. As the polarizers are crossed at maximum extinction, we can measure the extinction ratio $\sigma^2$
of the polarizers by measuring the following ratio:
\begin{eqnarray}
\sigma^2 =
\frac{I_{\mathrm{e}}}{I_{\mathrm{t}}}\Big|_{\Gamma=0}.
\end{eqnarray}
The value of $\sigma^2$ is regularly measured, in particular
before each shot for the Cotton-Mouton effect measurements. This
extinction ratio can typically vary from $4\times 10^{-7}$ to
$8\times 10^{-7}$.

As shown in Ref.\,\cite{Battesti2008}, because of the ellipticity
noise, the optical sensitivity improves when $\Gamma$ decreases.
Starting from $\Gamma = 0$ and rotating M$_1$ in the clockwise or
counterclockwise direction, we choose the sign of $\Gamma$ as well
as its value, with typically $\Gamma^2 \sim \sigma^2$. The sign of
$\Gamma$ is known by filling the vacuum chamber with nitrogen gas
and by measuring its Cotton-Mouton effect, whose sign and value
are perfectly known. This measurement has already been done with
this experiment and results are reported in
Ref.\,\cite{Berceau2012}. We performed several measurements with
different signs and values of $\Gamma$, showing that this
parameter is perfectly controlled. The value and the sign of
$\Gamma$ are set before each magnetic shot.

\begin{figure}[h]
\begin{center}
\includegraphics[width=9cm]{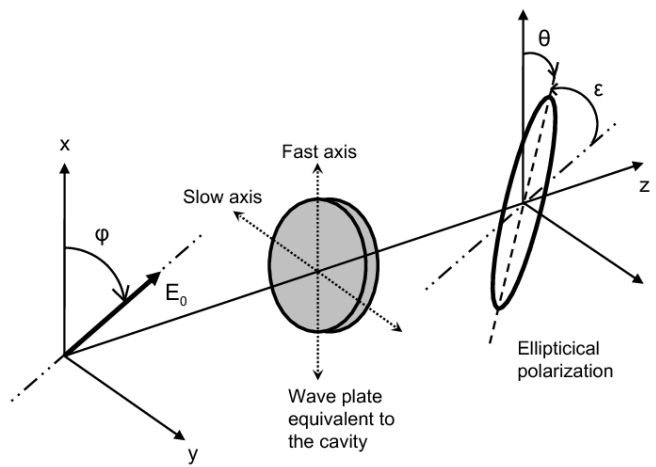}
\caption{\label{Fig:lame_eq_cavite} Rotation of the major axis of
the elliptical polarization due to the static birefringence of the
Fabry-P\'erot cavity.}
\end{center}
\end{figure}

The static birefringence of the cavity changes the incident linear
polarization into an elliptical polarization of ellipticity
$\Gamma$. But it also induces a rotation angle $\epsilon$ of the
major axis of the ellipse compared with the P polarizer axis. The
value of this angle can be calculated considering the Fabry-Pérot
cavity as an equivalent wave plate of phase retardation
$\delta_{\mathrm{eq}}$. The angle between the incident linear
polarization and the fast axis of the equivalent wave plate
corresponds to $\varphi$, as represented in
Fig.\,\ref{Fig:lame_eq_cavite}. The ellipticity induced by the
wave plate is given by:
\begin{eqnarray}
\Gamma = \frac{\sin(2\varphi)\sin(\delta_{\mathrm{eq}})}{2}.
\end{eqnarray}
As we set $\Gamma \ll 1$, the fast axis is almost aligned with $P$
and thus, we have $\varphi \ll 1$. Assuming that
$\delta_{\mathrm{eq}} \ll 1$, we get:
\begin{eqnarray}
\label{Eq:phi} \varphi = \frac{\Gamma}{\delta_{\mathrm{eq}}}.
\end{eqnarray}
We also have:
\begin{eqnarray}
\tan(2\theta) &=& \tan(2\varphi)\cos(\delta_{\mathrm{eq}}),\\
\theta &=& \varphi
\Big(1-\frac{\delta_{\mathrm{eq}}^2}{2}\Big),\label{Eq:theta}
\end{eqnarray}
where $\theta$ is the angle between the major axis of the ellipse
and the fast axis of the wave plate. Combining
Eqs.\,(\ref{Eq:phi}) and (\ref{Eq:theta}), we obtain the angle $\epsilon$ between the major axis of the elliptical polarization and the incident linear polarization:
\begin{eqnarray}
\epsilon = \theta - \varphi = -\frac{\Gamma
\delta_{\mathrm{eq}}}{2}. \label{Eq:epsilon}
\end{eqnarray}
The value of the phase retardation of our cavity is about
$|\delta_{\mathrm{eq}}| \sim 0.1$\,rad. This has been inferred by
measuring the value of $\Gamma$ as a function of the mirrors'
orientation, as explained in details in Ref.\,\cite{Bielsa2009}.
With a typical value of $|\Gamma|$ varying from $8 \times 10^{-4}$
to $3 \times 10^{-3}$, we obtain \mbox{$40~\mu\mathrm{rad}<|\epsilon|
<150~\mu$rad}.

\subsubsection{Cavity finesse and cavity filtering}\label{Par:filtering}

The finesse of the cavity is inferred from the measurement of the
photon lifetime $\tau$ inside the cavity. At $t=t_0$ the intensity
of the laser, previously locked onto the cavity resonance, is
switched off. The exponential decay of the intensity of the
ordinary beam for $t>t_0$ is fitted with:
\begin{eqnarray}
I_{\mathrm{t}}(t) =
I_{\mathrm{t}}(t_{\mathrm{0}})e^{-(t-t_{\mathrm{0}})/\tau},
\end{eqnarray}
to obtain $\tau$. The cavity finesse is related to the photon
lifetime through:
\begin{eqnarray}
F = \frac{\pi c \tau}{L_{\mathrm{c}}}.
\end{eqnarray}

The value of the photon lifetime is regularly checked during data
taking. In this experiment, it ranges from 1.06\,ms to 1.12\,ms,
corresponding to a finesse of 438\,000 to 465\,000. During a run
of data taking, the relative variation of the photon lifetime does
not exceed 2$\%$ at 1$\sigma$ confidence level.

Due to the photon lifetime, the cavity acts as a first order low
pass filter, as explained in details in Ref.\,\cite{Berceau2010}.
Its complex response function $H(\nu)$ is given by:
\begin{eqnarray}
H(\nu)=\frac{1}{1+i\frac{\nu}{\nu_{\mathrm{c}}}},\label{Eq:H_nu}
\end{eqnarray}
with $\nu$ the frequency and $\nu_{\mathrm{c}} = 1/4\pi\tau \simeq 70$\,Hz the cavity cutoff frequency. This filtering has to be taken into
account in particular for the time dependent magnetic field
applied inside the Fabry-Pérot cavity.

The cavity also acts as a first order low pass filter for the
ordinary beam $I_{\mathrm{t}}(t)$ compared to the beam incident on the
cavity. But, due to the cavity birefringence, the cavity acts as a
second order low pass filter for the extraordinary beam
$I_{\mathrm{e}}(t)$. This effect is explained in details in
Ref.\,\cite{Berceau2010}. The second order low pass filter
represents the combined action of two successive identical first order low pass filters. Their complex response function is given by Eq.\,(\ref{Eq:H_nu}). While the first one characterizes the usual
cavity behavior, we can interpret the second filter in terms of
pumping or filling: due to the mirror birefringence, some photons
of the ordinary beam are gradually converted into the
extraordinary beam at each reflection. Thus, if we want to
directly compare $I_{\mathrm{t}}(t)$ and $I_{\mathrm{e}}(t)$,
one has to apply the first order low pass filter to
$I_{\mathrm{t}}(t)$. The filtered signal $I_{\mathrm{t,f}}(t)$ is
then used for the analysis.

\subsection{Signals}\label{Par:signals}

The ellipticity $\Psi(t)$ and the rotation of the polarization
$\Theta_{\mathrm{F}}(t)$ induced by the transverse and the
longitudinal magnetic fields can be related to the ratio of the
extraordinary and ordinary powers as follows:
\begin{eqnarray}\label{Eq:Ie_It}
\frac{I_{\mathrm{e}}(t)}{I_{\mathrm{t,f}}(t)}=\sigma^2 + [\Gamma
+ \Psi(t)]^2 + [\epsilon + \Theta_{\mathrm{F}}(t)]^2.
\end{eqnarray}
This formula, which can be obtained using the Jones formalism,
clearly shows that our experiment is sensitive to both
ellipticities and rotations.

\section{Faraday effect of helium gas}

As stated above, the Faraday effect corresponds to a magnetic
circular birefringence $\Delta n_\mathrm{F}$ induced by a
longitudinal magnetic field $B_\|$. Form Eqs.\,(\ref{Eq:kF}),
(\ref{Eq:ThetaF_intro}) and (\ref{Eq:Theta_theta}), we deduce that
the polarization rotation to be measured depends on $k_\mathrm{F}$
as follows:
\begin{equation}
\Theta_\mathrm{F}(t) =
2F\frac{L_B}{\lambda}k_\mathrm{F}{B_\|}(t).\label{Eq:Theta_F_general}
\end{equation}
For historical reasons, Faraday effect is usually given in terms
of the Verdet constant $V$\,\cite{Verdet1854}, that is related to
the Faraday constant as:
\begin{eqnarray}
V = \frac{\pi k_\mathrm{F}}{\lambda}.\label{Eq:k_F}
\end{eqnarray}
Eq.\,(\ref{Eq:Theta_F_general}) becomes:
\begin{equation}
\Theta_\mathrm{F}(t) =
\frac{2F}{\pi}VB_\|(t)L_B.\label{Eq:Verdet_Finesse}
\end{equation}

\subsection{Magnetic field}\label{Par:longitudinal_field}

\begin{figure}
\begin{center}
\includegraphics[width=8cm]{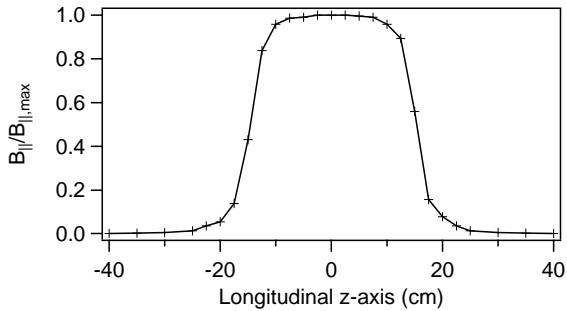}
\caption{\label{Fig:B_profil_long.eps} Normalized profile of the
longitudinal magnetic field inside the solenoid along the
longitudinal $z$-axis. The crosses correspond to the
measurements.}
\end{center}
\end{figure}

To measure the Faraday effect, we need a longitudinal magnetic
field. It is delivered thanks to a 300\,mm long solenoid. Its
diameter is 50\,mm and it corresponds to 340 loops of copper wire.
The magnetic field profile along the longitudinal $z$-axis has
been measured with a gaussmeter. Fig.\,\ref{Fig:B_profil_long.eps}
shows the normalized profile. We define $B_{\|,0}$ as the maximum
magnetic field, thus at the center of the coil, and $L_B$ as the
equivalent magnetic length such that:
\begin{equation}
\int^{+\infty}_{-\infty} B_\|(z)dz \equiv B_{\|,0}L_B.
\end{equation}
This equivalent magnetic length has been calculated by numerically
integrating the measured field. Taking into account the
experimental uncertainties, we obtain $L_B = (0.308 \pm 0.006)$\,m
at 1$\sigma$. We can reach a maximum magnetic field of about
$4.3$\,mT corresponding to an injected current of 3\,A.

To measure the magnetic field during operation, we measure the
current injected into the coil. The form factor $B_\|/I$ has been
determined experimentally using the gaussemeter and an ammeter.
This form factor remains constant for frequency modulation ranging
from DC to 50\,Hz. Finally we have estimated the relative
uncertainty $u(B_\|)/B_\|=1.4\%$ at 1$\sigma$, taking into account
the uncertainties coming from the gaussmeter, the ammeter and from a
possible small misalignment of the laser beam inside the solenoid.

Faraday effect measurements were performed at room temperature $T
= 293$\,K in an air-conditioned room. When a current is injected
into the solenoid, the temperature increases inside the coil.
Nevertheless, for a maximum current of 3\,A, the increase is lower
than 2\,K. This will be taken into account in the final
uncertainty.

\subsection{Analysis of Faraday signal}

The magnetic field at the center of the coil is modulated at the
frequency $\nu$: $B_\| = B_{\|,0}\sin(2\pi \nu t + \phi)$. The
rotation of the polarization due to the Faraday effect is thus
given by:
\begin{eqnarray}
\Theta_{\mathrm{F}} &=& \Theta_0\sin(2\pi \nu t +\phi),\\
\mathrm{with}\; \Theta_0 &=& \frac{2F}{\pi} V B_{\|,0} L_B.
\label{Eq:Theta_0}
\end{eqnarray}

Expanding Eq.\,(\ref{Eq:Ie_It}), we obtain:
\begin{equation}
\frac{I_{\mathrm{e}}(t)}{I_\mathrm{t,f}(t)} = \sigma^2 +
\epsilon^2 +2\epsilon\Theta_\mathrm{F}(t) + \Theta_\mathrm{F}^2(t) +
\Gamma^2 + 2\Gamma\Psi(t) + \Psi^2(t).\label{Eq:Iext_It_general_Faraday}
\end{equation}
We define the ratio between the Faraday and the Cotton-Mouton
signals as:
\begin{equation}
R_\mathrm{F/CM} =
\frac{2\epsilon\Theta_\mathrm{F}+\Theta_\mathrm{F}^2} {2\Gamma
\Psi + \Psi^2}.\label{Eq:R_F_CM}
\end{equation}
For the Faraday measurements, our typical static ellipticity is
$\Gamma = 3\times 10^{-3}$\,rad, and Eq.\,(\ref{Eq:epsilon}) gives
$|\epsilon| = 150\,\mu$rad. We evaluate the value of
$R_\mathrm{F/CM}$ using the theoretical values of the Verdet and
Cotton-Mouton constants of helium which are given later in this
article. For this experiment, our typical helium pressure is
$30\times 10^{-3}$\,atm and the cavity finesse is of the order of
465\,000 corresponding to a cavity cutoff frequency of about
70\,Hz. The solenoid mainly induces a longitudinal magnetic field,
but, for the sake of argument, let's perform the calculation with
the same value 4.3\,mT for the longitudinal and the transverse
magnetic field. One gets: $R_\mathrm{F/CM} \sim 10^6$. The
Cotton-Mouton effect is thus negligible.
Eq.\,(\ref{Eq:Iext_It_general_Faraday}) thus becomes:
\begin{equation}
\frac{I_{\mathrm{e}}(t)}{I_{\mathrm{t,f}}(t)}=\sigma^2 +
\Gamma^2 + [\epsilon +
\Theta_{\mathrm{F}}(t)]^2.\label{Eq:Ie_It_Faraday}
\end{equation}

This equation results in three main frequency components:
\begin{eqnarray}
I_\mathrm{DC} &=& \sigma^2+\Gamma^2+\epsilon^2+\frac{\Theta_0^2}{2},\\
I_\mathrm{\nu} &=& \frac{2\epsilon\Theta_0}
{\sqrt{1+\big(\frac{\nu}{\nu_\mathrm{c}}\big)^2}} \sin\Big[2\pi
\nu t +\phi -
\arctan\Big(\frac{\nu}{\nu_\mathrm{c}}\Big)\Big],\label{Eq:Faraday_f}\\
I_\mathrm{2\nu} &=& -\frac{\Theta_0^2}
{2\sqrt{1+\big(\frac{2\nu}{\nu_\mathrm{c}}\big)^2}} \cos\Big[4\pi
\nu t +2\phi -
\arctan\Big(\frac{2\nu}{\nu_\mathrm{c}}\Big)\Big].\nonumber\\
\label{Eq:Faraday_2f}
\end{eqnarray}
As mentionned before, the cavity acts as a first-order low-pass filter,
with a cavity cutoff frequency $\nu_\mathrm{c}$. This filtering
has been taken into account in Eqs.\,(\ref{Eq:Faraday_f}) and
(\ref{Eq:Faraday_2f}).

The amplitude of the $\nu$-component, $I_\nu$ depends on
$\Theta_0$ but also on $\epsilon$ whose value is not precisely
known. On the other hand, $I_\mathrm{2\nu}$ only depends on
$\Theta_0$. Consequently it is the only component used to measure
the Verdet constant. The amplitude of the $2\nu$ frequency
component, proportional to $(B_{\|,0} L_B)^2$, is measured as a
function of the magnetic field amplitude. We fit our data by $K_V
B_{\|,0}^2$. The Verdet constant $V$ finally depends on the
measured experimental parameters as follows, using
Eq.\,(\ref{Eq:Theta_0}) and the amplitude of the $2 \nu$-component given in Eq.(\ref{Eq:Faraday_2f}):
\begin{equation}
V(T,P) = \sqrt{\frac{K_V}{2}}
\frac{\Big[1+{(8\pi\tau\nu)}^{2}\Big]^{1/4}}
{2\tau\Delta^{\mathrm{FSR}}L_B}.\label{Eq:Verdet}
\end{equation}

\subsection{Results}

\subsubsection{Our result}

\begin{figure}
\begin{center}
\includegraphics[width=8cm]{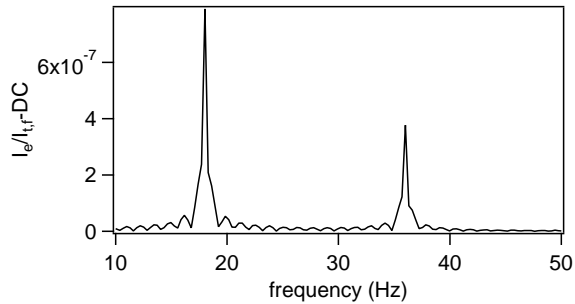}
\caption{\label{Fig:FFT_Faraday} Fourier transform of
$I_\mathrm{e}/I_\mathrm{t}-DC$ with about $60\times 10^{-3}$\,atm
of helium and with $B_{\|,0} L_B = 1.3\times 10^{-3}$\,T.m. The
magnetic frequency modulation is $\nu = 18$\,Hz.}
\end{center}
\end{figure}

We report in Fig.\,\ref{Fig:FFT_Faraday} the Fourier transform of
\mbox{$I_{\mathrm{e}}/I_\mathrm{t,f}-DC$-signal} with about $60\times 10^{-3}$\,
atm of helium and with $B_{\|,0} L_B = 1.3\times 10^{-3}$\,Tm. The
magnetic frequency modulation is fixed to $\nu = 18$\,Hz in order
to have the $2\nu$ frequency lower than the cavity cutoff
frequency. We can observe both components at frequencies $\nu$ and
$2\nu$. During the Faraday data taking, the photon lifetime was
$\tau = (1.12\pm 0.02)$\,ms corresponding to a cavity finesse of
($465\,000 \pm 8\,000)$.

\begin{figure}
\begin{center}
\includegraphics[width=8cm]{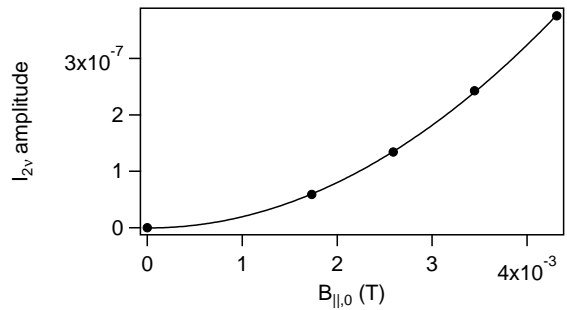}
\caption{\label{Fig:amplitude_vs_B0} Amplitude of the $2\nu$
frequency component as a fonction of $B_{\|,0}$ with about
$60\times 10^{-3}$\, atm of helium. The solid line corresponds to
the quadratic fit of the experimental data.}
\end{center}
\end{figure}

We plot in Fig.\,\ref{Fig:amplitude_vs_B0} the amplitude of the $2\nu$ component as a function of $B_{\|,0}$. We fit our data by
a quadratic law $K_V B_{\|,0}^2$. We also study the $\nu$
frequency component as a function of the magnetic field amplitude.
According to the relation (\ref{Eq:Faraday_f}), we obtain a linear
dependance. By fitting these data by a linear equation and using
the value of the Verdet constant measured with the $2\nu$
frequency component, we infer the value of the $\epsilon$
parameter. We obtain $\epsilon \simeq 10^{-4}$\,rad, in agreement
with the value calculated with Eq.\,(\ref{Eq:epsilon}).

\begin{figure}
\begin{center}
\includegraphics[width=8cm]{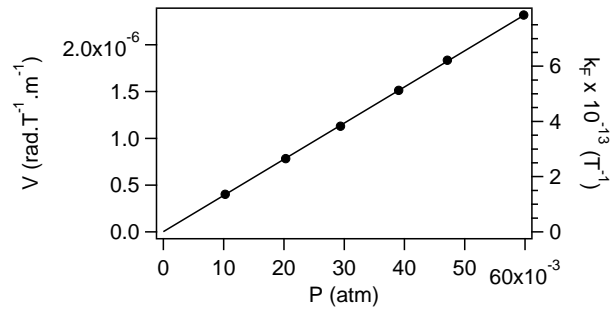}
\caption{\label{Fig:Verdet_vs_P} Verdet constant of helium as a
function of pressure. The solid line corresponds to the linear fit
of the experimental data.}
\end{center}
\end{figure}

We performed Faraday constant measurements at different pressures
from $10^{-2}$ to $6 \times 10^{-2}$\,atm. They are summarized in
Fig.\,\ref{Fig:Verdet_vs_P}. We measure the gas pressure in the
chamber with pressure gauges which have a relative uncertainty
given by the manufacturer of 0.2$\%$. In this range of pressure,
helium can be considered as an ideal gas and the pressure
dependence of the Verdet constant is thus linear. As shown in
Fig.\,\ref{Fig:Verdet_vs_P}, our data are correctly fitted by a
linear equation. Its $V$-axis intercept is consistent with zero
within the uncertainties. Its slope gives the normalized Verdet
constant at $\lambda = 1064$\,nm and at $T = (294\pm1)$\,K:
\begin{equation}\label{Eq:V_293K}
V = (3.87 \pm 0.12)\times 10^{-5}\; \mathrm{atm}^{-1}\mathrm{rad.
T}^{-1}\mathrm{m}^{-1}.
\end{equation}
With a scale law on the gas density and considering an ideal gas,
this corresponds to a normalized Verdet constant at $T =
273.15$\,K of:
\begin{equation}
V = (4.17 \pm 0.13)\times 10^{-5}\; \mathrm{atm}^{-1}\mathrm{rad.
T}^{-1}\mathrm{m}^{-1}.
\end{equation}
The uncertainty is given at 1$\sigma$. It is calculated from the
relative A and B-type uncertainties summarized in
Table\,\ref{Tab:TableBudget} and detailed in
Ref.\,\cite{Berceau2012}. Using Eq.\,(\ref{Eq:k_F}), we can also
give the normalized Faraday constant. At $T = 273.15$\,K, one
gets:
\begin{equation}
k_\mathrm{F} = (1.41 \pm 0.04)\times 10^{-11}\;
\mathrm{atm}^{-1}\mathrm{T}^{-1}.
\end{equation}

\begin{table}
\begin{center}
\begin{tabular*}{0.47\textwidth}{p{2.5cm}  p{3cm}  p{3cm}}
  \hline
  \hline
  \centering Parameter &  Relative A-type & Relative B-type\\
  & uncertainty & uncertainty\\
 \hline
 \centering $\tau$         &  \raggedright $2\times 10^{-2}$   & \\
 \centering $K_V$            &  $8\times 10^{-3}$ & \\
 \centering $B_{\|,0}$          &  &  $1.4\times 10^{-2}$   \\
 \centering $L_B$        &   &  $2.0\times 10^{-2}$   \\
 \centering $\Delta^{\mathrm{FSR}}$   &  &  $3\times 10^{-4}$   \\
 \centering $P$   &  &  $2\times 10^{-3}$  \\
 \hline
 \hline
\end{tabular*}
\end{center}
\caption{\label{Tab:TableBudget} Parameters and their respective
relative A and B-type uncertainties at 1$\sigma$ that have to be
measured to infer the value of the normalized Verdet constant
$V$.}
\end{table}

\subsubsection{Comparison}

\begin{figure}
\begin{center}
\includegraphics[width=8cm]{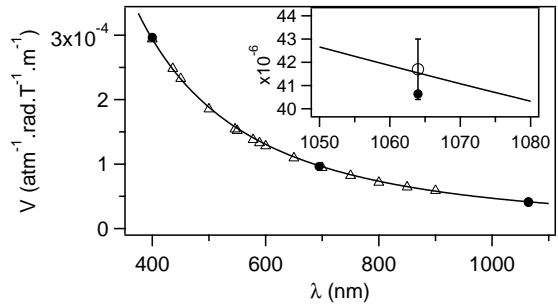}
\caption{\label{Fig:Comparaison_Faraday} $\vartriangle$:
Experimental values of helium normalized Verdet constant at $T =
273.15$\,K reported in Ref.\,\cite{Ingersoll1956} for wavelength
from 363\,nm to 900\,nm. These values are fitted by the law
$A/\lambda^2$ (solid line). $\circ$: Our experimental value at $T
= 273.15$\,K. $\bullet$: Theoretical predictions at $T =
273.15$\,K reported in Ref.\,\cite{Ekstrom2005}. Inset : Zoom
around $\lambda = 1064$\,nm. The error bar corresponds to the
1$\sigma$ uncertainty of our measurement.}
\end{center}
\end{figure}

Our value of the normalized Verdet constant can be compared to
other published values. Ref.\,\cite{Ingersoll1956} presents the
most extensive experimental values in helium. They have been
measured at different wavelengths, from 363\,nm to 900\,nm, and
they correspond to the open triangles in
Fig.\,\ref{Fig:Comparaison_Faraday} at $T = 273.15$\,K. As stated
by the authors in Refs.\,\cite{Ingersoll1956,Ingersoll1954}, ``the
average absolute probable error is considered to be about 1$\%$'',
but ``the scale of measurement was determined by a comparison of
these results with accepted values for water''. This is an
important difference from our experiment since we do not need to
calibrate our setup with another gas. All parameters from which
the measured Verdet constant depends are accurately monitored,
yielding therefore a Verdet constant of high precision and
accuracy.

As far as we know, no value has been reported at 1064\,nm, our
working wavelength. Nevertheless, it can be quadratically
interpolated from the data of Ref.\,\cite{Ingersoll1956} with a
fit $A/\lambda^2$ (solid line in
Fig.\,\ref{Fig:Comparaison_Faraday}). This gives a normalized
Verdet constant at $\lambda = 1064$\,nm and $T = 273.15$\,K of $V
= (4.15\pm 0.05)\times 10^{-5}\,\mathrm{atm}^{-1}\mathrm{rad.
T}^{-1}\mathrm{m}^{-1}$. The uncertainty is the one given by the
fit. This value is compatible with ours, which is represented as
the open circle in the inset of
Fig.\,\ref{Fig:Comparaison_Faraday}.

We finally compared our value with the theoretical predictions at
$T = 273.15$\,K. The most recent ones were published in
2005\,\cite{Ekstrom2005} exploiting a four-component Hartree-Fock
calculations and in 2012\,\cite{Savukov2012} using a relativistic
particle hole configuration interaction method.
Ref.\,\cite{Ekstrom2005} gives values at different wavelengths that
are plotted in Fig.\,\ref{Fig:Comparaison_Faraday} with the filled
points. The value at $\lambda = 1064$\,nm is $V = 4.06\times
10^{-5}\,\mathrm{atm}^{-1}\mathrm{rad. T}^{-1}\mathrm{m}^{-1}$ and
it is plotted in Fig.\,\ref{Fig:Comparaison_Faraday} with the
filled point. Ref.\,\cite{Savukov2012} does not give a value at
1064\,nm, but it can be obtained by a quadratic interpolation
the data provided by the authors. One obtains $V = (4.09\pm
0.02)\times
10^{-5}\,\mathrm{atm}^{-1}\mathrm{rad.T}^{-1}\mathrm{m}^{-1}$,
with an uncertainty given by the fit. Both theoretical values are
compatible with our experimental Verdet constant. All these
theoretical and experimental values are summarized in
Table\,\ref{Tab:ComparaisonFaraday}.

\begin{table}
\begin{center}
\begin{tabular*}{0.47\textwidth}{p{1.5cm}  p{3cm}  p{4cm}}
  \hline
  \hline
  Ref. & \centering$V\times 10^{5}$  & Remarks\\
  & \centering[$\mathrm{atm}^{-1}\mathrm{rad.
T}^{-1}\mathrm{m}^{-1}$] & \\
 \hline
 & Theoretical values & \\
  & & \\
 \cite{Ekstrom2005} & \centering4.06 & \\
 \cite{Savukov2012} &  \centering$4.09\pm 0.02$ & quadratically interpolated\\
 \hline
 & Experimental values & \\
 & & \\
 \multirow{3}{*}{\cite{Ingersoll1956}} & \centering\multirow{3}{*}{$4.15 \pm 0.05$} &  quadratically interpolated   \\
 & &  not absolute: scaled to\tabularnewline
 & & water\tabularnewline
 This work & \centering$4.17 \pm 0.13$ & \\
 \hline
 \hline
\end{tabular*}
\end{center}
\caption{\label{Tab:ComparaisonFaraday} Experimental and
theoretical values of the normalized Verdet constant at $T =
273.15$\,K, $\lambda = 1064$\,nm and with uncertainties at 1$\sigma$.}
\end{table}

\section{Cotton-Mouton effect of helium gas}

The Cotton-Mouton effect consists in a linear birefringence
$\Delta n_\mathrm{CM}$ induced by a transverse magnetic field
$B_\perp$. From Eqs. (\ref{Eq:Psi_intro}) and (\ref{Eq:Psi_psi})
we deduce that the ellipticity $\Psi(t)$ to be measured is linked
to $k_\mathrm{CM}$ by:
\begin{eqnarray}
\Psi(t) &=& 2F\frac{L_B}{\lambda}k_{\mathrm{CM}}B_\perp^2(t)
\sin2\theta_\mathrm{P}.
\end{eqnarray}
The angle $\theta_\mathrm{P}$ is adjusted to 45 degrees with the
experimental procedure explained in Ref.\,\cite{Berceau2012}.

\subsection{Magnetic field}\label{Par:transverse_field}

One can see that $\Psi$ is proportional to $B_{\perp}^2 L_B$. In
order to have $\Psi$ as high as possible, we have to maximize this
parameter. This is fulfilled using pulsed fields delivered by one
magnet, named X-coil, especially designed by the LNCMI. The
principle of this magnet and its properties are described in
details in Refs. \cite{Battesti2008, Batut2008}. It can provide a
maximum field of more than 14T over an equivalent length $L_B$ of
0.137\,m \cite{Berceau2012}. The high voltage connections can be
remotely switched to reverse the direction of the field. Thus we
can set $\textbf{B}$ parallel or anti-parallel to the
$x$-direction, as shown in Fig.\,\ref{Fig:ExpSetup}.

The pulsed coil is immersed in a liquid nitrogen cryostat to limit
its heating. A pause between two pulses is necessary
to let the magnet cool down to the equilibrium temperature. We do
not need to use the coil at its maximum field since the
sensitivity of our experiment is largely sufficient. We have chosen to apply a maximum field of 3\,T in order to limit the ageing of
the magnet. From one shot to another, a relative variation of the
maximum of the field lower than 1.5\,$\%$ was observed due to a
variation of the power supply voltage.

\begin{figure}[h]
\begin{center}
\includegraphics[width=8cm]{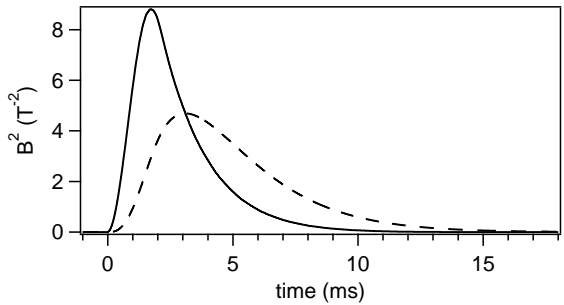}
\caption{\label{Fig:B}  Square of the magnetic field amplitude as
a function of time for a maximum field of 3\,T. Solid black curve:
$B_{\perp}^2$, dashed curve: $B^2_{\perp,\mathrm{f}}$.}
\end{center}
\end{figure}

The pulse duration is less than 10\,ms, with the maximum of the
field reached within 2\,ms. Since the pulse duration is of the
same order of magnitude as the photon lifetime inside the cavity,
the filtering of the Fabry-Pérot cavity has to be taken into
account for the magnetic field, as said in
section\,\ref{Par:filtering}. We calculate the filtered field
$B^2_{\perp,\mathrm{f}}$ from $B_{\perp}^2$ by using the
first-order low pass filter corresponding to the cavity. The time
profiles of $B_{\perp}^2$ and $B^2_{\perp,\mathrm{f}}$ are shown
in Fig.\,\ref{Fig:B}, for a maximum field of 3\,T.

\subsection{Analysis of Cotton-Mouton signal}

As mentioned in section \ref{Par:signals}, the ratio of power
$I_{\mathrm{e}}$ and $I_\mathrm{t,f}$ is linked to the
birefringence $\Psi(t)$ to be measured as follows:
\begin{eqnarray}
\frac{I_{\mathrm{e}}(t)}{I_{\mathrm{t,f}}(t)}=\sigma^2 + [\Gamma
+ \Psi(t)]^2 + [\epsilon + \Theta_{\mathrm{F}}(t)]^2,
\label{Eq:Ie_It_bis}
\end{eqnarray}
where $\Theta_{\mathrm{F}}(t)$ is the rotation angle due to a
longitudinal component of the pulsed magnetic field inducing a
Faraday effect in helium. This component $B_\parallel(t)$ is
firstly due to the X-structure of the coil. It is around
230\,times smaller than the transverse field, $i.e.$ around 10\,mT
for a pulse of 3\,T. Moreover a contribution to $B_\parallel$
appears if the cryostat is not perfectly aligned with the optical
axis. The diameter of the cryostat is 60\,cm. A typical
misalignment of 2\,mm over this length, i.e. around 3\,mrad, leads
to a longitudinal component of 10\,mT. Finally the estimated
longitudinal magnetic field is about 20\,mT. It can be present during
a shot over an equivalent length $L_B = 0.137$\,m.

Using Eq.\,(\ref{Eq:Verdet_Finesse}) and the value of the Verdet
constant given in Eq.\,(\ref{Eq:V_293K}), we can calculate the
rotation of the polarization $\Theta_{\mathrm{F}}$ due to
$B_{\parallel}$. It is about 30\,mrad per atmosphere of helium
gas. We then calculate the ratio of the Faraday effect over
the Cotton-Mouton effect $R_{\mathrm{F/CM}}$, given by
Eq.\,(\ref{Eq:R_F_CM}). Since the static ellipticity is
typically $|\Gamma| \simeq 8 \times 10^{-4}$\,rad corresponding to
$\epsilon \simeq 40\,\mu$rad as stated in section
\ref{Par:bir_cavite}, this ratio goes from 200 at
$40\times10^{-3}$\,atm to 2600 at $550\times10^{-3}$\,atm. This
shows that the Faraday effect component is not negligible and thus
need to be taken into account.

From Eq.\,(\ref{Eq:Ie_It_bis}), we obtain:
\begin{equation}
\frac{I_{\mathrm{e}}(t)}{I_{\mathrm{t,f}}(t)} = \sigma^2 + \Gamma^2 + \epsilon^2 + 2\Gamma\Psi(t) + \Psi^2(t)
 + 2\epsilon\Theta_{\mathrm{F}}(t) + \Theta_{\mathrm{F}}^2(t).
\end{equation}
This formula shows that the angle $\epsilon$ carries the Faraday
effect of the gas. During a Cotton-Mouton effect measurement we
want to have the Faraday effect as small as possible. We therefore
minimize $\epsilon$ before the shot, once the value of $\Gamma$ is
set, by turning the analyzer A. As we can see in Fig.
\ref{Fig:lame_eq_cavite}, it consists in aligning A, which has
been initially adjusted at 90 degrees compared to the incident
polarization, on the minor axis of the elliptical polarization.
Nevertheless, in order to take into account the imperfections of
this experimental adjustment, we still keep $\epsilon$ in the
formula, assuming that $\epsilon^2 \ll \Gamma^2$.

To extract the ellipticity $\Psi(t)$, we calculate the following
$Y(t)$ function:
\begin{eqnarray}\label{Eq:Y(t)}
Y(t) &=& \frac{\frac{I_{\mathrm{e}}(t)}{I_{\mathrm{t,f}}(t)} - DC}{2|\Gamma|}\\
\nonumber &=& \gamma \Psi(t) + \frac{\Psi^2(t)}{2|\Gamma|}+\gamma \frac{|\epsilon|\Theta_{\mathrm{F}}(t)}{2|\Gamma|} + \frac{\Theta_{\mathrm{F}}^2(t)}{2|\Gamma|},
\end{eqnarray}
where $\gamma$ corresponds to the sign of $\Gamma$. $DC$ is the
static signal:
\begin{eqnarray}
DC = \sigma^2 + \Gamma^2 + \epsilon^2 = \Big<
\frac{I_{\mathrm{e}}(t)}{I_{\mathrm{t,f}}(t)}\Big>_{t<0},
\end{eqnarray}
and it is measured just before each shot, the magnetic field being
applied at $t = 0$. We also measure the extinction ratio $\sigma^2$ before each shot using the experimental procedure
described in section\,\ref{Par:bir_cavite}. The absolute value of
the static ellipticity is then calculated as follows:
\begin{eqnarray}
|\Gamma| = \sqrt{\Big<
\frac{I_{\mathrm{e}}(t)}{I_{\mathrm{t,f}}(t)}\Big>_{t<0} -
\sigma^2}.
\end{eqnarray}

Two parameters are adjustable in the experiment: the sign $\gamma$ of the static ellipticity $\Gamma$ and the direction of the transverse
magnetic field. We acquire signals for both signs of $\Gamma$ and
both directions of $\textbf{B}$: parallel to $x$ is
denoted as $>0$ and antiparallel is denoted as $<0$. This gives
four data series: ($\Gamma>0$, $B_{\perp}>0$), ($\Gamma>0$,
$B_{\perp}<0$), ($\Gamma<0$, $B_{\perp}<0$) and ($\Gamma<0$,
$B_{\perp}>0$). For each series, signals calculated with
Eq.\,(\ref{Eq:Y(t)}) are averaged and denoted as $Y_{>>}$,
$Y_{><}$, $Y_{<<}$ and $Y_{<>}$. The first subscript corresponds
to $\Gamma
>0$ or $<0$ while the second one corresponds to
$\textbf{B}$ parallel or antiparallel to $x$.

The $Y$ signals are the sum of different effects with different
symmetries, denoted as $S$:
\begin{eqnarray}
\nonumber Y_{>>} &=& a_{>>}S_{++} + b_{>>}S_{+-} + c_{>>}S_{--} + d_{>>}S_{-+},\\
\nonumber Y_{><} &=& a_{><}S_{++} - b_{><}S_{+-} - c_{><}S_{--} + d_{><}S_{-+},\\
\nonumber Y_{<<} &=& a_{<<}S_{++} - b_{<<}S_{+-} + c_{<<}S_{--} - d_{<<}S_{-+},\\
\nonumber Y_{<>} &=& a_{<>}S_{++} + b_{<>}S_{+-} - c_{<>}S_{--} -
d_{<>}S_{-+}.\\
\label{Eq:Y}
\end{eqnarray}
The first subscript in $S$ corresponds to the symmetry towards the
sign of $\Gamma$ and the second one towards the direction of
$\textbf{B}$. The subscript $+$ indicates an even parity
while the subscript $-$ indicates an odd parity. In practice
$w_{>>} \simeq w_{><} \simeq w_{<<} \simeq w_{<>}$ (with $w =
a,~b,~c$ or $d$) depend on the experimental parameters. These
values are not perfectly equal because the experimental parameters
slightly vary from one shot to another, in particular the value of
$|\Gamma|$.

Possible physical effects contributing to the different $S$
signals are summarized in Tab. \ref{Tab:S_effects}. The $S_{+-}$
signal does not appear in Eq.\,(\ref{Eq:Y(t)}) but it has to be
taken into account. It corresponds to a signal with an odd parity
towards the direction of $\textbf{B}$ and an even parity towards
the sign of $\Gamma$ that could be, for example, a spurious effect
on the photodiodes Ph$_{\mathrm{t}}$ and Ph$_{\mathrm{e}}$ induced
by the magnetic field.

Linear combinations of the $Y$ signals allow to highlight the
effect corresponding to the different symmetries:
\begin{eqnarray}
\nonumber J_1 &=& \frac{Y_{>>} + Y_{><} + Y_{<<} + Y_{<>}}{4},\\
\nonumber &=& \overline{a}~S_{++} + \Delta b_1~S_{+-} + \Delta c_1~S_{--} + \Delta d_1~S_{-+},\\
\nonumber J_2 &=& \frac{Y_{>>} - Y_{><} - Y_{<<} + Y_{<>}}{4},\\
\nonumber&=& \overline{b}~S_{+-}+ \Delta a_2~S_{++} + \Delta c_2~S_{--} + \Delta d_2~S_{-+},\\
\nonumber J_3 &=& \frac{Y_{>>} - Y_{><} + Y_{<<} - Y_{<>}}{4},\\
\nonumber&=& \overline{c}~S_{--}+ \Delta a_3~S_{++} + \Delta b_3~S_{+-} + \Delta d_3~S_{-+},\\
\nonumber J_4 &=& \frac{Y_{>>} + Y_{><} - Y_{<<} - Y_{<>}}{4},\\
\nonumber &=& \overline{d}~S_{-+} + \Delta a_4~S_{++} + \Delta b_4~S_{+-} + \Delta c_4~S_{--}.\\
\end{eqnarray}
with $\Delta w_i \simeq 0$ ($w = a,~b,~c$ or $d$ and $i = 1,~2,~3$
or $4$). The signal we want to measure is $\Psi(t)$ which
corresponds to the main part of $S_{-+}(t)$, thus proportional to
$B^2_{\perp,\mathrm{f}}$. We can write:
\begin{eqnarray}\label{Eq:J4}
\nonumber J_4 &=& \alpha B^2_{\perp,\mathrm{f}} + \Delta a_4~S_{++} + \Delta b_4~S_{+-} + \Delta c_4~S_{--},\\
&\simeq& \alpha B^2_{\perp,\mathrm{f}}.
\end{eqnarray}
We fit the function $J_4$ by $\alpha
B^2_{\perp,\mathrm{f}}$ to obtain $\alpha$. The Cotton-Mouton
constant $k_{\mathrm{CM}}$ finally depends on the measured
experimental parameters as follows:
\begin{eqnarray}\label{Eq:kcm}
k_{\mathrm{CM}}(T,P) = \frac{\alpha}{4\pi \tau
\Delta^\mathrm{FSR}}\frac{\lambda}{L_B}\frac{1}{\sin2\theta_\mathrm{P}}.
\end{eqnarray}
The terms $T$ and $P$ correspond to the gas temperature and
pressure.

\begin{center}
\begin{table}[h]
\begin{center}
\begin{tabular}{m{2cm} m{4cm}}
\hline \hline \centering $S$ signal &  \centering Physical effect
\tabularnewline \hline \centering $S_{++}(t)$ & \centering
$\Theta_{\mathrm{F}}^2(t)$, $\Psi^2(t)$ \tabularnewline \centering
$S_{+-}(t)$ & \centering B effects on photodiodes...
\tabularnewline \centering $S_{--}(t)$ & \centering
$\gamma\Theta_{\mathrm{F}}(t)$ \tabularnewline \centering
$S_{-+}(t)$ & \centering $\gamma\Psi(t)$\tabularnewline \hline
\hline
\end{tabular}
\end{center}
\caption{Possible physical effects contributing to the $S$
signals.} \label{Tab:S_effects}
\end{table}
\end{center}

\subsection{Results}

\subsubsection{Our result}

We have taken data for helium pressures ranging from
$40\times10^{-3}$\,atm to $550\times10^{-3}$\,atm. Before
injecting the gas, we pumped the vacuum chamber and the initial
pressure was about $10^{-10}$\,atm. Several series of four shots
($\Gamma>0$, $B_{\perp}>0$; $\Gamma>0$, $B_{\perp}<0$; $\Gamma<0$,
$B_{\perp}<0$ and $\Gamma<0$, $B_{\perp}>0$) have been acquired
for each pressure. The vacuum chamber was pumped between two
measurements at different pressures, which made them totally
independent. The temperature of the gas during the magnetic pulse
was measured previously \cite{Berceau2012} and was $T = (293 \pm
1)$\,K. For this set of measurement the mean photon lifetime
inside the cavity is $\tau = (1.06 \pm 0.02)$\,ms, corresponding
to a finesse of $438\,000\pm 8\,000$.

\begin{center}
\begin{figure}[htp!!]
\centering
\subfloat[$Y_{>>}$]{\includegraphics[width=4.2cm]{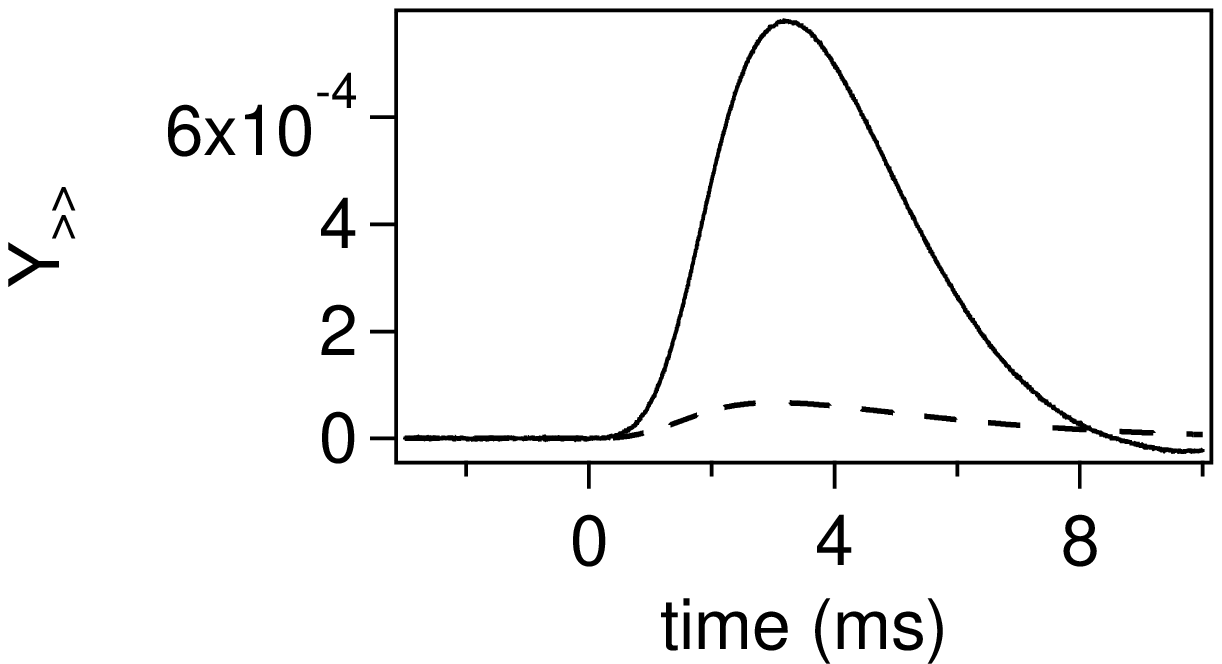}}
\subfloat[$Y_{><}$]{\includegraphics[width=4.2cm]{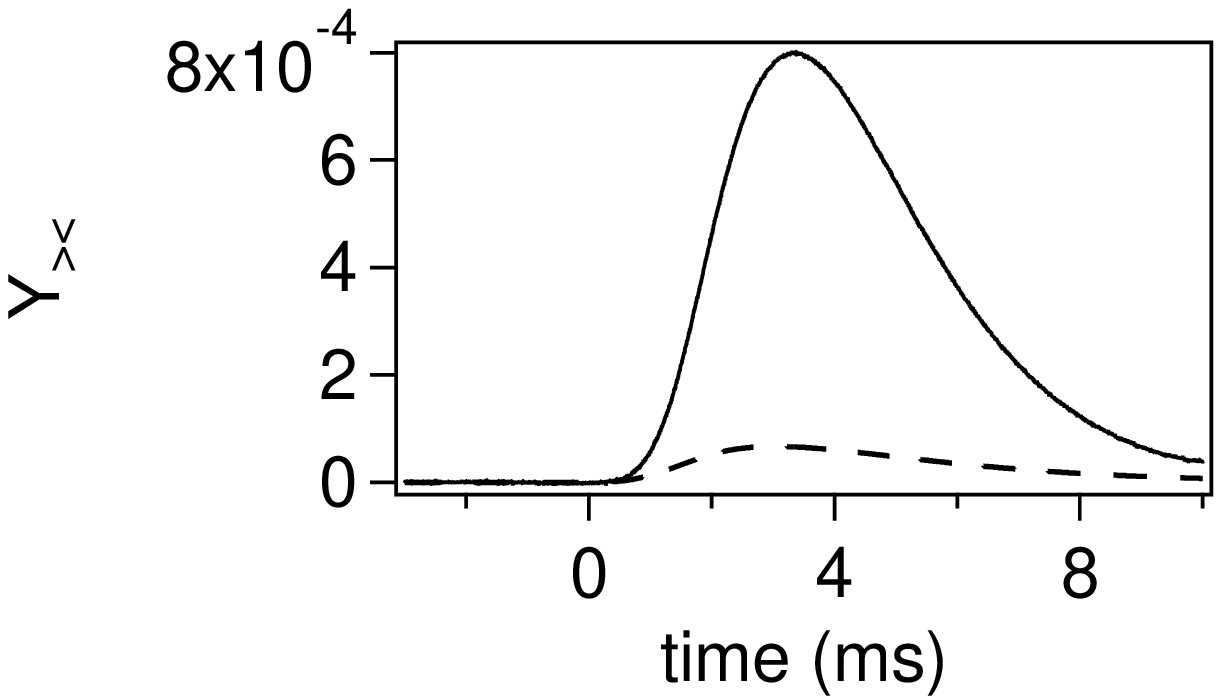}}\\
\subfloat[$Y_{<<}$]{\includegraphics[width=4.2cm]{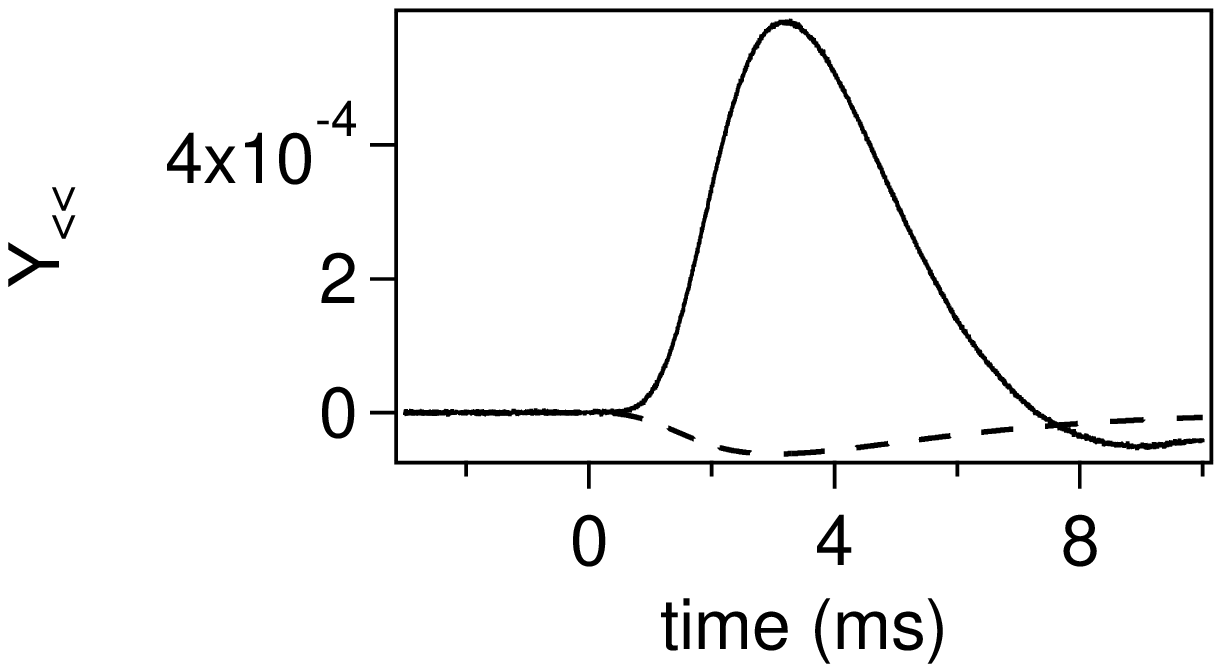}}
\subfloat[$Y_{<>}$]{\includegraphics[width=4.2cm]{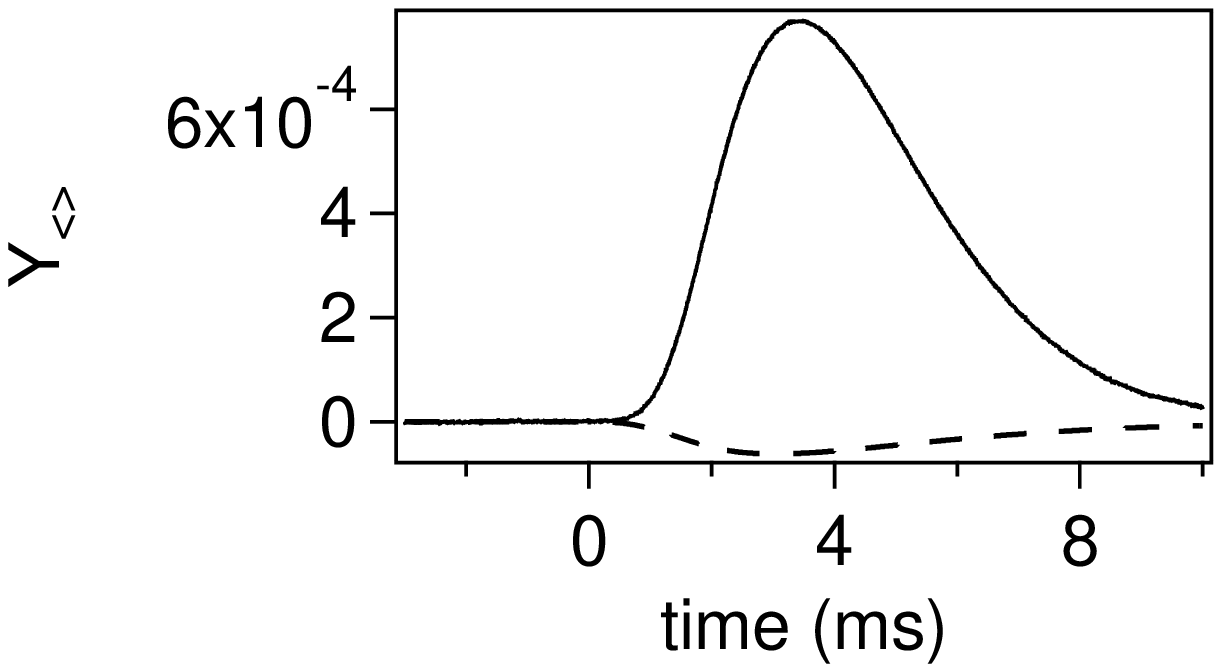}}\\
\caption{\small{Time evolution of the $Y(t)$ signals at a pressure
of $550\times10^{-3}$\,atm. Solid black curve: experimental data,
dashed curve: expected signal from the theoretical prediction
considering only the Cotton-Mouton effect.}} \label{Fig:Y_550}
\end{figure}
\end{center}

The signals $Y_{>>}$, $Y_{><}$, $Y_{<<}$ and $Y_{<>}$ obtained for
a pressure of $550 \times 10^{-3}$\,atm are plotted in Fig.\,\ref{Fig:Y_550}. We
calculate the signals expected from the theoretical prediction
considering only the Cotton-Mouton effect \cite{Rizzo2005}. The
theoretical signals (dashed line) are superimposed to the
experimental data (solid line). One can see that the $Y$ signals
do not match at all with the expected signals. A more refined study is thus
needed to extract the Cotton-Mouton effect. The $Y$ signals are in
fact linear combinations of different effects with different
symmetries towards the sign of $\Gamma$ and the direction of
$\textbf{B}$, as predicted in Eqs.\,(\ref{Eq:Y}).

\begin{center}
\begin{figure}[htp!!]
\centering \subfloat[$J_1$]{\includegraphics[width=4.2cm]{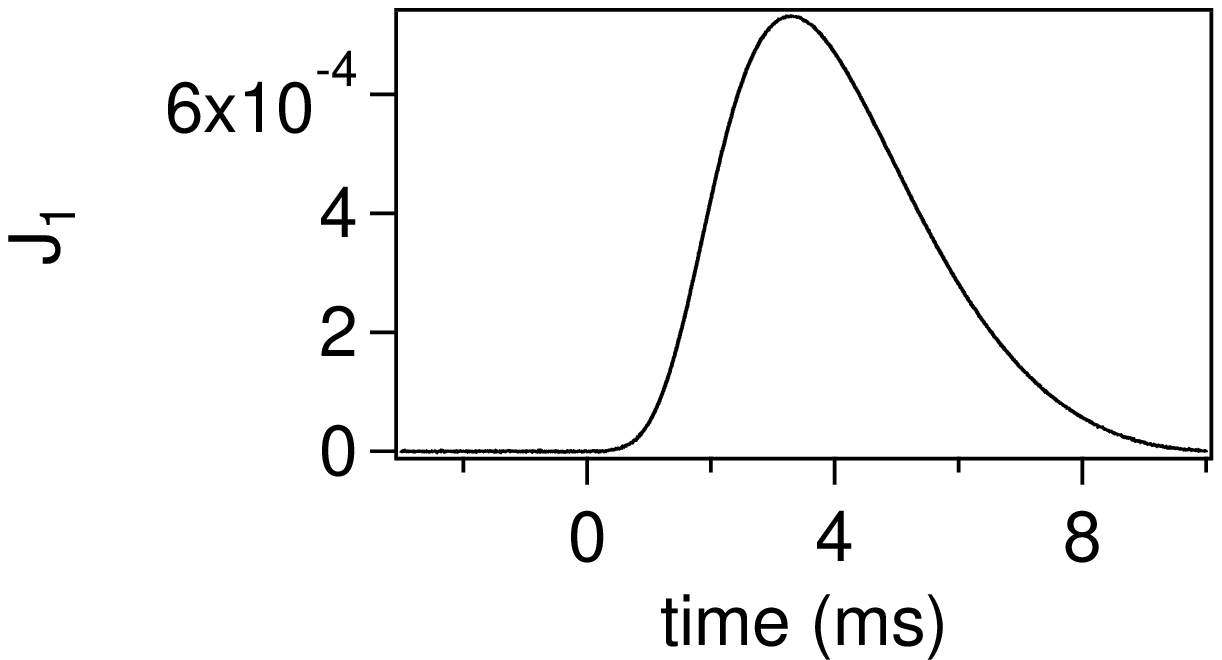}}
\subfloat[$J_2$]{\includegraphics[width=4.2cm]{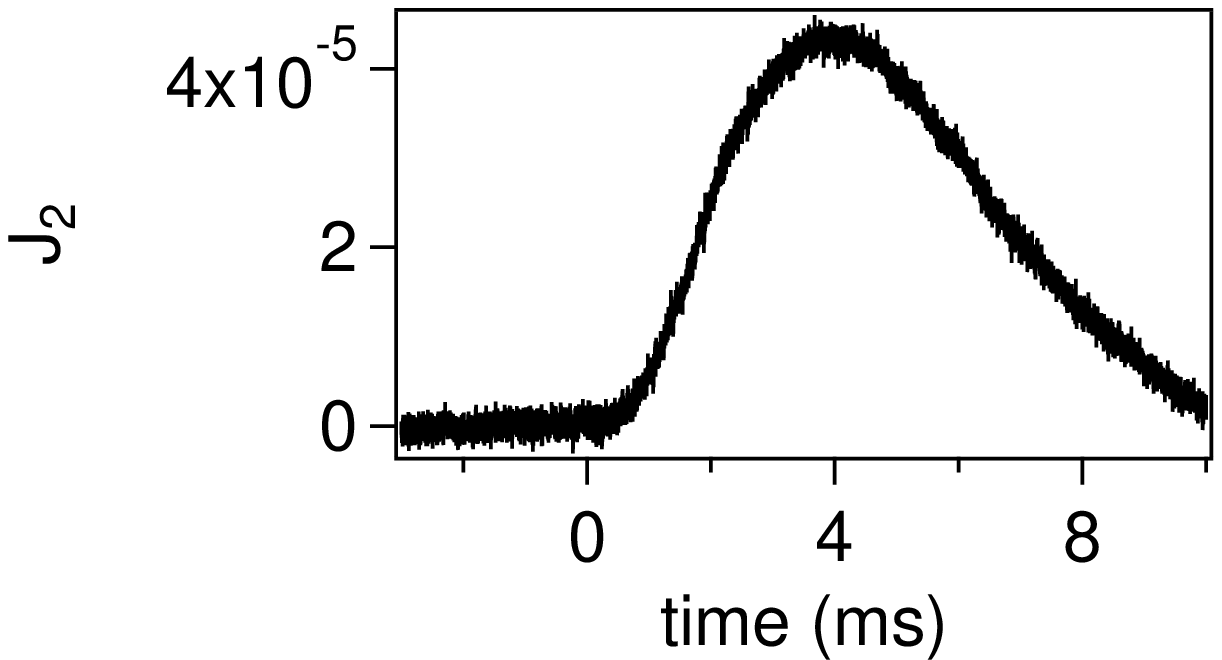}}\\
\subfloat[$J_3$]{\includegraphics[width=4.2cm]{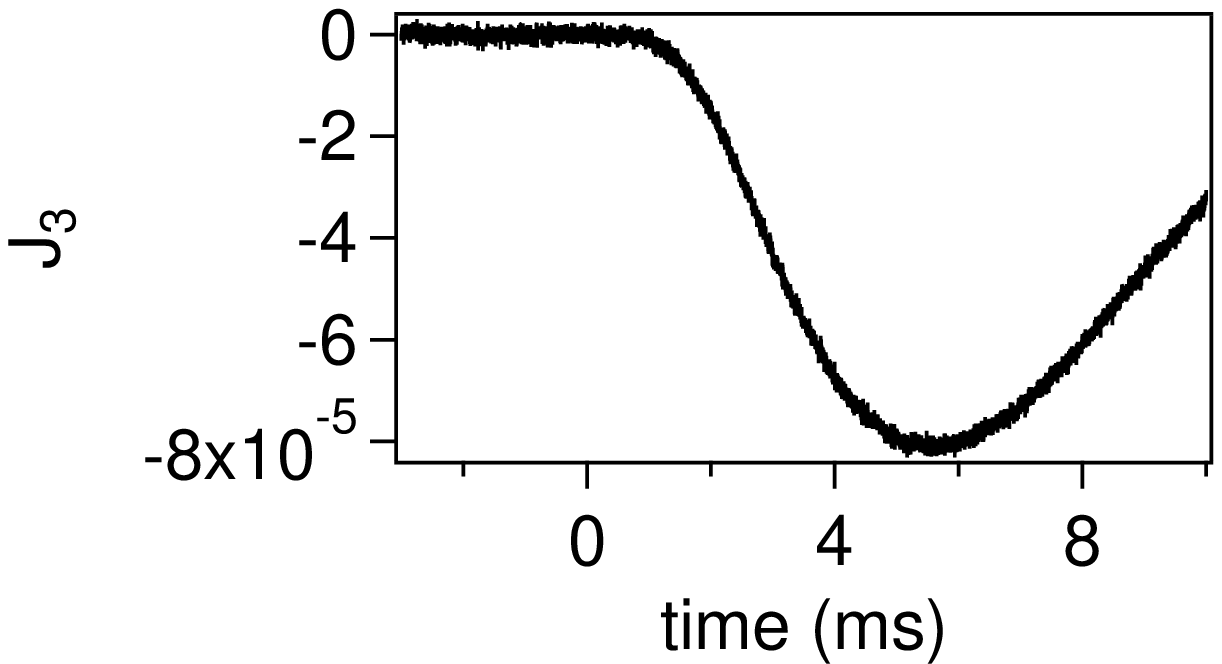}}
\subfloat[$J_4$]{\includegraphics[width=4.2cm]{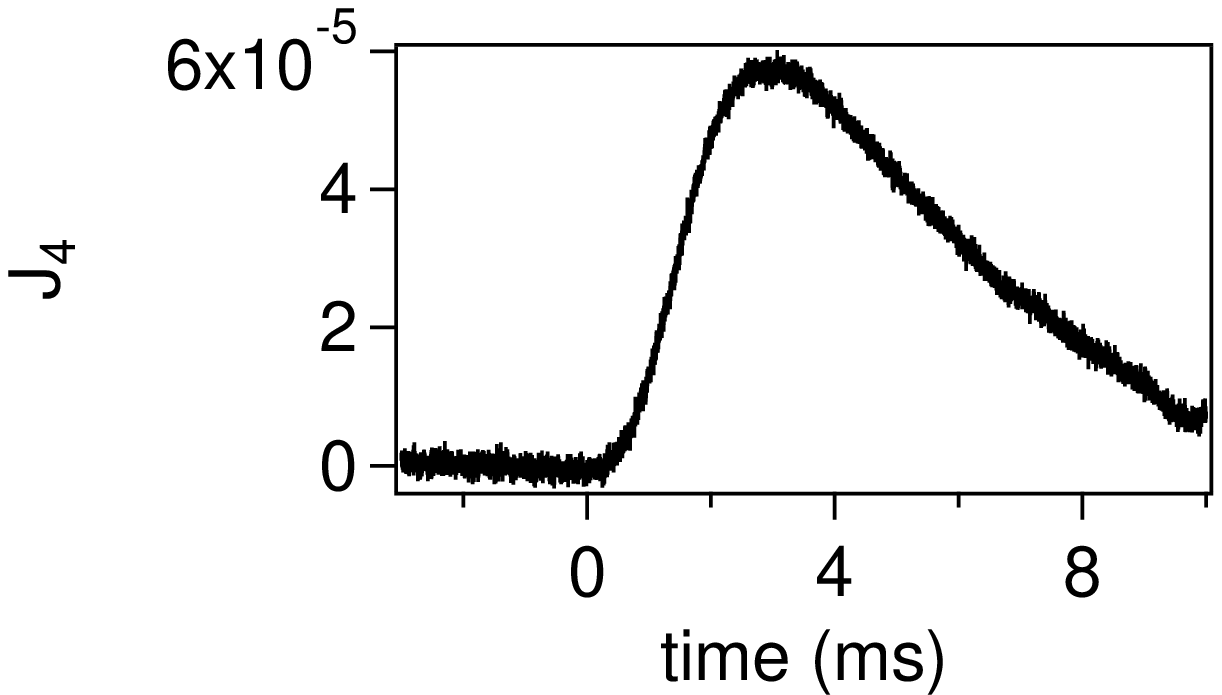}}\\
\caption{\small{Time evolution of the $J(t)$ signals at a pressure
of $550\times10^{-3}$\,atm.}} \label{Fig:J_550}
\end{figure}
\end{center}

\begin{center}
\begin{figure}[htp!!]
\centering \subfloat[Maximum of $J_1$. Dashed curve: quadratic
fit]{\includegraphics[width=4.2cm]{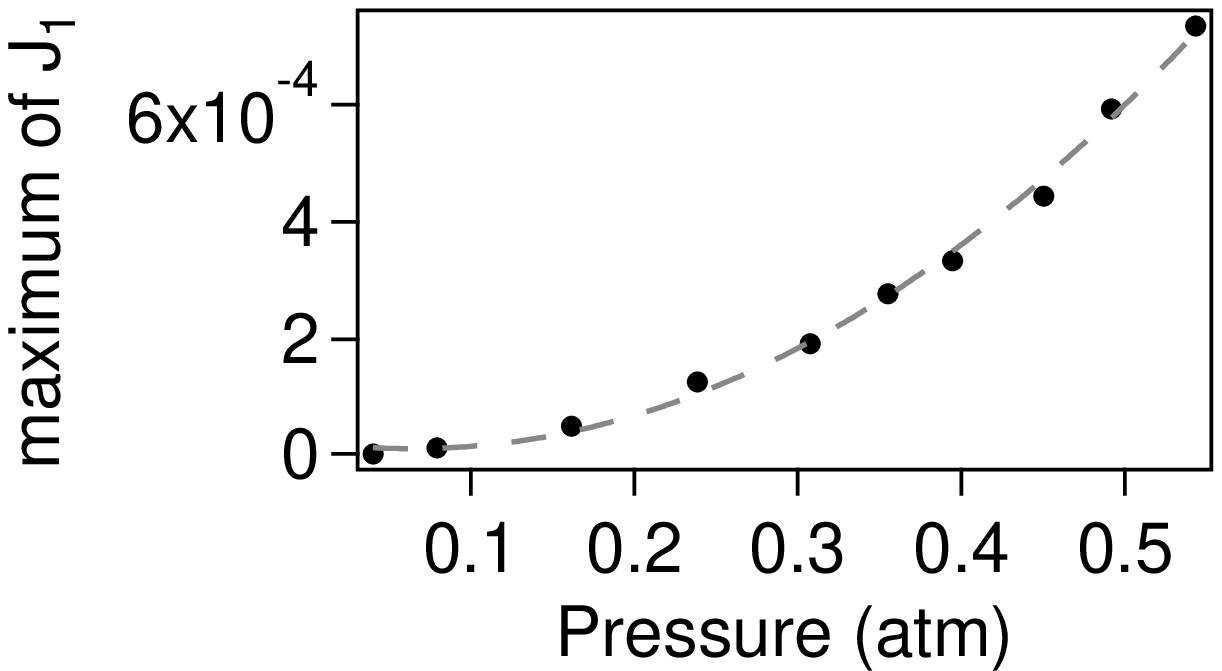}}
\subfloat[Maximum of $J_2$. Dashed curves: linear and quadratic fits]{\includegraphics[width=4.2cm]{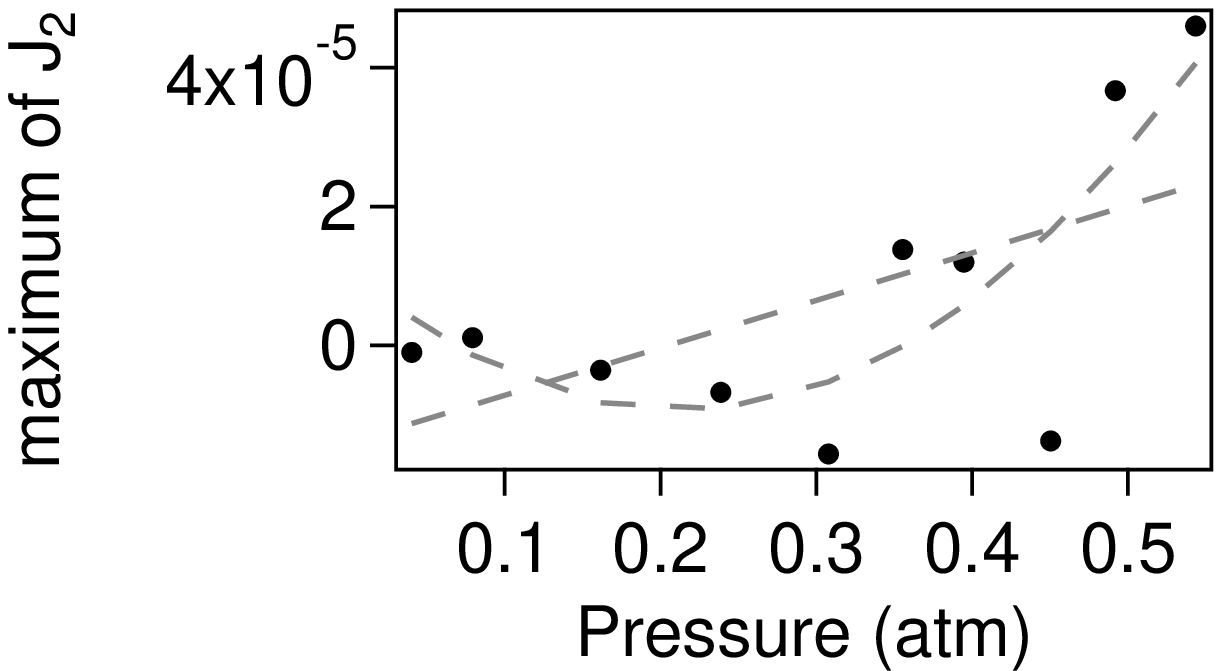}}\\
\subfloat[Maximum of $J_3$. Dashed curve: linear
fit]{\includegraphics[width=4.2cm]{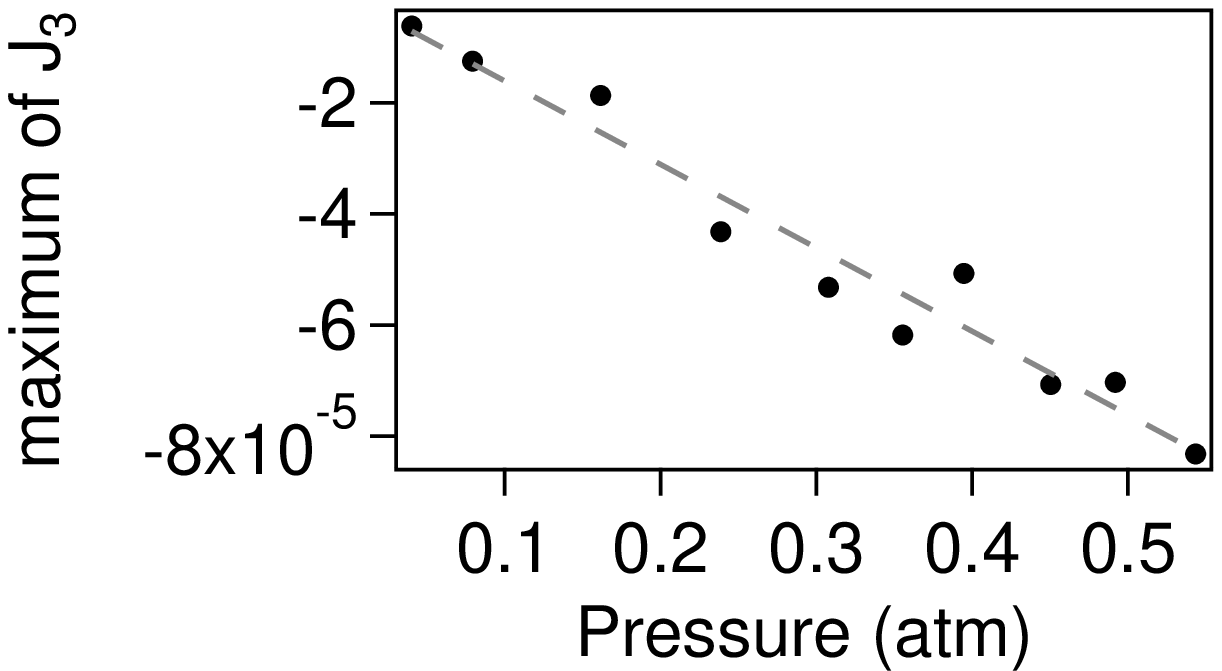}}
\subfloat[Maximum of $J_4$. Dashed curve: linear fit]{\includegraphics[width=4.2cm]{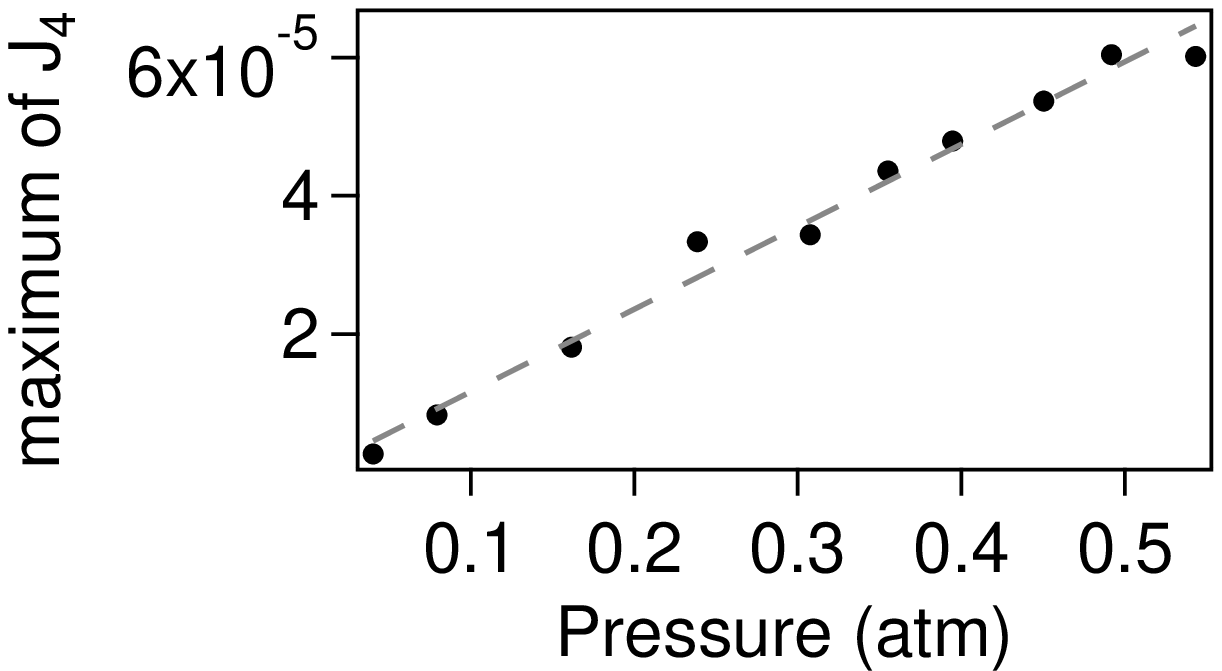}}\\
\caption{\small{Evolution of the maximum of the $J(t)$ signals as
a function of pressure.}} \label{Fig:max_vs_P}
\end{figure}
\end{center}

We then calculate the corresponding $J$ signals, plotted in
Fig.\,\ref{Fig:J_550}. In order to validate the physical origin of
$J_1$, $J_2$, $J_3$ and $J_4$, we have studied the evolution of
the value of their maximum as a function of pressure. They are
shown in Fig.\,\ref{Fig:max_vs_P}. In this range of pressure,
helium can be considered as an ideal gas and the pressure
dependence of the Faraday and Cotton-Mouton effects is thus
linear. We see that the maxima of $J_3$ and $J_4$ are
proportional to the pressure, which is consistent with the Faraday
effect due to the residual longitudinal magnetic field
$B_\parallel$ and the Cotton-Mouton effect due to the transverse
magnetic field $B_\perp$. The maximum of $J_1$ increases with the
square of the pressure. It confirms that this signal contains the
terms $\Theta_{\mathrm{F}}^2$ and $\Psi^2$. The value of the $J_2$
maximum does not have a clear dependence with the pressure.
Moreover the shape of $J_2(t)$ is not the same from a pressure to
another. Finally, the $J_2$ signals can be fitted by a linear
combination of $J_1$, $J_3$ and $J_4$. Thus, we deduce that $J_2$
is essentially a linear combination of the other signals, and that
the signal $\overline{b}S_{+-}$ is almost zero.


Thus we can write:
\begin{eqnarray}
\nonumber J_1 &\simeq& \overline{a}~S_{++},\\
\nonumber J_2 &\simeq& \Delta a_2~S_{++} + \Delta c_2~S_{--} + \Delta d_2~S_{-+},\\
\nonumber J_3 &\simeq& \overline{c}~S_{--},\\
J_4 &\simeq& \overline{d}~S_{-+}.
\end{eqnarray}
The main contribution to $J_4$ comes from the Cotton-Mouton
effect. We thus fit $J_4(t)$ with $\alpha
B^2_{_{\perp},\mathrm{f}}(t)$. The value of $k_{\mathrm{CM}}$ is
then calculated thanks to Eq.\,(\ref{Eq:kcm}).

\begin{figure}[h]
\begin{center}
\includegraphics[width=8cm]{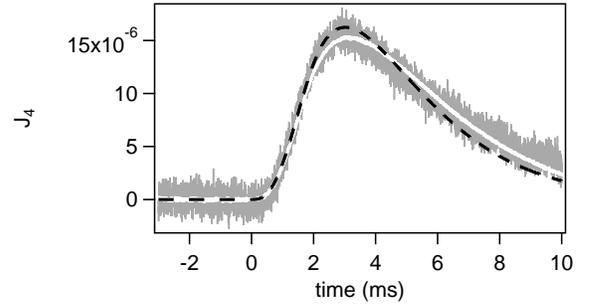}
\caption{\label{Fig:fit_J4_162} \small{Gray: Time evolution of
$J_4$ for a pressure of $162\times10^{-3}$\,atm. Black dashed
curve: fit with $\alpha B^2_{\perp,\mathrm{f}}$. White solid
curve: fit with $\alpha B^2_{\perp,\mathrm{f}} + \alpha_1 J_1 +
\alpha_3 J_3$, the value of $\alpha$ being fixed at the value
obtained with the previous fit $\alpha B^2_{\perp,\mathrm{f}}$.}}
\end{center}
\end{figure}

For the lowest pressures, the Cotton-Mouton signal, proportional
to $\alpha B^2_{\perp,\mathrm{f}}$, also decreases. In this case,
$\Delta a_4\,S_{++}$ and $\Delta c_4\,S_{--}$ are not completely
negligible compared to $\alpha B^2_{\perp,\mathrm{f}}$. This is
shown in Fig.\,\ref{Fig:fit_J4_162} where a typical signal
obtained for a helium pressure of $162\times10^{-3}$\,atm is
plotted. We see that the fit of $J_4$ with $\alpha
B^2_{\perp,\mathrm{f}}$ does not perfectly match the experimental
data. To obtain a better fit, we have to add parameters. To this
end, we first fix the value of $\alpha$ at the value obtained with
the first fit $\alpha B^2_{\perp,\mathrm{f}}$. Then we fit $J_4$
with $\alpha B^2_{\perp,\mathrm{f}} + \alpha_1 J_1 + \alpha_3
J_3$. $J_2$ is not used in this fit because, as said before, it is
mainly a linear combination of the other signals. One can see in
Fig.\,\ref{Fig:fit_J4_162} that this fit now matches much better the data. We can conclude that, in this case, we have:
\begin{eqnarray}
J_4 &=& \alpha B^2_{\perp,\mathrm{f}} + \Delta a_4~S_{++} + \Delta
c_4~S_{--},\\
&=& \alpha B^2_{\perp,\mathrm{f}} + \frac{\Delta
a_4}{\overline{a}}~J_1 + \frac{\Delta c_4}{\overline{c}}~J_3,
\end{eqnarray}
with $\alpha_2 = \Delta a_4/\overline{a}$ and $\alpha_3 = \Delta
c_4/\overline{c}$. This fit procedure repeated for each pressure
shows that we always have $\alpha_2$ and $\alpha_3$ lower than
0.1.

\begin{figure}[h]
\begin{center}
\includegraphics[width=8cm]{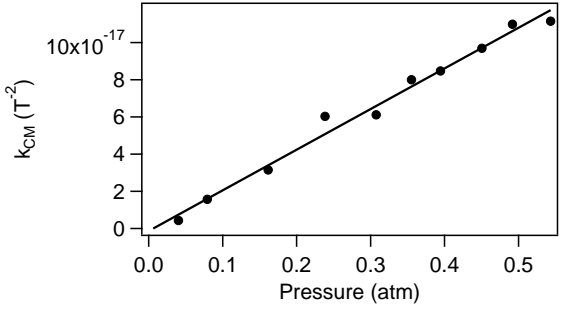}
\caption{\label{Fig:kcm_vs_P} Linear magnetic birefringence of
helium gas as a function of pressure. The solid line corresponds
to the linear fit of the experimental data.}
\end{center}
\end{figure}

The value of $k_{\mathrm{CM}}$ as a function of the pressure is
shown in Fig.\,\ref{Fig:kcm_vs_P}. A linear fit of this data gives
$k_{\mathrm{CM}} = (2.19 \pm 0.09) \times 10^{-16}$
T$^{-2}$atm$^{-1}$ at $T =(293 \pm 1)$K. Its
$k_{\mathrm{CM}}$-axis intercept is consistent with zero within
the uncertainties.

The A-type uncertainties come from the fit and from the photon
lifetime with a relative variation lower than 2$\%$. The B-type
uncertainties have been evaluated previously and detailed in
Ref.\,\cite{Berceau2012}. They essentially come from the length of
the magnetic field $L_B$. They are summarized in Table
\ref{Tab:Err_CM}. We obtain for the value of the Cotton-Mouton
constant at $T = (293 \pm 1)$\,K:
\begin{eqnarray}
k_{\mathrm{CM}} = (2.19 \pm 0.12)\times
10^{-16}\,\mathrm{T^{-2}atm^{-1}}.
\end{eqnarray}
The value of $k_{\mathrm{CM}}$ normalized at 273.15\,K is
calculated with a scale law on the gas density:
\begin{eqnarray}
k_{\mathrm{CM}} = (2.35 \pm 0.13)\times
10^{-16}\,\mathrm{T^{-2}atm^{-1}},
\end{eqnarray}
at $\lambda = 1064$\,nm, taking into account the uncertainty on
the temperature.

\begin{table}
\begin{center}
\begin{tabular*}{0.47\textwidth}{m{2cm}  m{3cm}  m{3cm}}
  \hline
  \hline
 \centering \multirow{2}{*}{Parameter} &   \multirow{2}{*}{Typical value} &  Relative B-type\\
 & & uncertainty\\
 \hline
\centering$\alpha$         &   $10^{-5}$\,rad\,T$^{-2}$     &  $2.2 \times 10^{-2}$ \\
 \centering$\Delta^\mathrm{FSR}$    &   65.996\,MHz &  $3 \times 10^{-4}$   \\
 \centering$L_{B}$          &   0.137\,m &  $2.2 \times 10^{-2}$   \\
 \centering$\lambda$        &   1064.0\,nm &  $<5 \times 10^{-4}$   \\
\centering$\sin 2\theta_\mathrm{P}$   &   1.0000 &  $9 \times 10^{-4}$   \\
 \hline
\centering Total & & $3.1 \times 10^{-2}$\\
 \hline
 \hline
\end{tabular*}
\end{center}
\caption{\label{Tab:Err_CM} Parameters that have to be measured to infer the value of the Cotton-Mouton constant $k_{\mathrm{CM}}$ and their respective relative B-type uncertainty at 1$\sigma$.}
\end{table}

\subsubsection{Comparison}

The value of the Cotton-Mouton effect in helium is calculated very
precisely thanks to {\it ab initio} quantum chemistry
computational methods \cite{Rizzo2005}. Theoreticians concentrate
on the calculation of the hypermagnetizability anisotropy $\Delta
\eta$ while experimentalists measure the birefringence $\Delta
n_{\mathrm{CM}} = k_{\mathrm{CM}} B^2$. The Cotton-Mouton
constant $k_{\mathrm{CM}}$ is linked to $\Delta \eta$ by
\cite{Rizzo1997}:
\begin{eqnarray}\label{Eq:delta_eta}
k_{\mathrm{CM}}\ [\mathrm{atm^{-1}T^{-2}}] = \frac{6.18381 \times
10^{-14}}{T} \Delta \eta\ [\mathrm{a.u.}]
\end{eqnarray}

Few experiments were realized to measure the Cotton-Mouton
effect of helium. The results are summarized in Table
\ref{Tab:kcm_He}. The theoretical values correspond to the ones of
Ref.\,\cite{Coriani1999}. The latter have been obtained using the
Full Configuration Interaction (FCI) method and the most extended
wave functions basis. They are expected therefore to be very
accurate.
\begin{center}
\begin{table}[h]
\begin{center}
\begin{tabular}{m{1.5cm} m{1.1cm} m{2.8cm} | m{2.8cm}}
\hline \hline \multicolumn{3}{c|}{Experimental results} &
\centering Theoretical prediction
\cite{Coriani1999}\tabularnewline \hline \centering \multirow{2}{*}{Ref.}&
\centering\multirow{2}{*}{$\lambda$ [nm]} & \centering\multirow{2}{*}{$10^{16} \times
k_{\mathrm{CM}}~[T^{-2}]$}  &\centering
\multirow{2}{*}{$10^{16} \times k_{\mathrm{CM}}~[T^{-2}]$}
\tabularnewline
 & & & \tabularnewline
\hline \centering\cite{Cameron1991} &\centering
$514.5$ &\centering $1.80 \pm 0.36$ & \centering2.3959
\tabularnewline \centering\cite{Bregant2009} & \centering$532$ &
\centering$2.08 \pm 0.16$ & \centering2.3966 \tabularnewline
\centering\cite{Muroo2003} & \centering$790$ &\centering $3.95
\pm 1.40$ & \centering2.4018 \tabularnewline
\centering\cite{Bregant2009} &\centering $1064$ &\centering $2.22
\pm 0.16$ & \centering2.4036 \tabularnewline
\centering This work &\centering $1064$ &\centering $2.35
\pm 0.13$ & \centering2.4036 \tabularnewline \hline \hline
\end{tabular}
\end{center}
\caption{Experimental and theoretical values of Cotton-Mouton constant for helium gas. Values are normalized for a temperature of 273.15 K and a pressure of 1 atm. Uncertainties are given at 1$\sigma$.}
\label{Tab:kcm_He}
\end{table}
\end{center}

\begin{center}
\begin{figure}[htp!!]
\centering
\subfloat[Reported values for $\lambda$ ranging from 514.5 nm to 1064 nm]{\includegraphics[width=8cm]{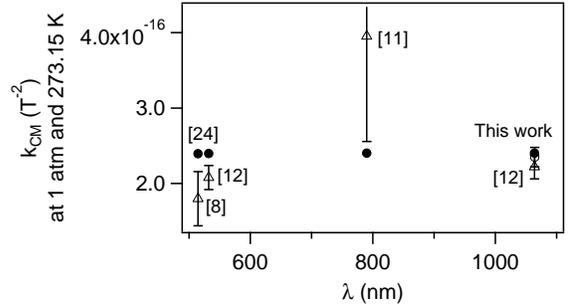}}\\
\subfloat[Summary of the two values at $\lambda = 1064$ nm]{\includegraphics[width=8cm]{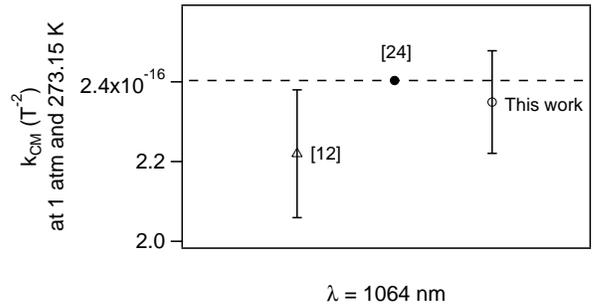}}\\
\caption{\small{Comparison of reported values of Cotton-Mouton
effect of helium gas. $\vartriangle$: experimental values of
helium Cotton-Mouton constant reported in
Refs.\,\cite{Cameron1991, Muroo2003, Bregant2009}. $\circ$: our
experimental value. $\bullet$ and dashed line: theoretical
predictions reported in Ref.\,\cite{Coriani1999}.}}
\label{Fig:kcm_biblio}
\end{figure}
\end{center}

Our result is compatible at better than 1$\sigma$ with the
theoretical prediction and is the most precise value of
$k_{\mathrm{CM}}$ ever measured, as we can see in
Fig.\,\ref{Fig:kcm_biblio} that summarizes the results for the
Cotton-Mouton measurements at 273.15\,K.

\section{Discussions and conclusion}

In this paper we report a new measurement of Faraday and
Cotton-Mouton effects at $\lambda = 1064$\,nm. Both measurements
have precisions that are of the order of a few percent,
corresponding to one of the most precise birefringence
measurements. Our measurements are also accurate and they are in
agreement with theory at better than 1$\sigma$.

As far as Faraday effect is concerned, our measurement is the first
at $\lambda = 1064$\,nm. It is worthwhile to stress that our
measurement is also absolute, while previous results
\cite{Ingersoll1954,Ingersoll1956} were given with respect to the
Faraday effect of water.

Our Cotton-Mouton measurement is the second experimental value at
$\lambda = 1064$\,nm but it is the first to agree so well with the
theoretical prediction. This definitely solves the problem of the
discrepancy between experiment and theory originated from the
first 1991 measurements and calculation \cite{Rizzo1997} and that
still persisted (see Table\,\ref{Tab:kcm_He}).

The measurement of such small Cotton-Mouton effects, as the helium
one, is not only important to test the quantum chemistry
predictions. It is also a crucial test for the apparata devoted to the
search of vacuum magnetic birefringence. Quantum electrodynamics
predicts that a vacuum, as any other centro symmetric medium,
should exhibit a Cotton-Mouton effect \cite{Battesti2013}. This
fundamental prediction has not yet been experimentally proven.
Several attempts have been made and a few are still under way
\cite{Battesti2013}. Vacuum Cotton-Mouton effect should be about
eight orders of magnitude smaller than the one of helium at
1\,atm. Measurement of the Cotton-Mouton effect of helium is
therefore compulsory in the search for improving the sensitivity
of such apparata.

Our experimental method based on pulsed fields coupled to a
Fabry-Pérot cavity seems very appropriate to reach the sensitivity
needed for vacuum measurement. The measurements reported here
validate the whole procedure of data taking and signal analysis
that allow to isolate the main effect from the spurious ones thanks to signal symmetries. They are therefore a milestone in the road
towards vacuum linear magnetic birefringence.

\begin{acknowledgments}

We thank all the members of the BMV collaboration, and in
particular J. B\'eard, J. Billette, P. Frings, B. Griffe, J.
Mauchain, M. Nardone, J.-P. Nicolin and G. Rikken for strong
support. We are also indebted to the whole technical staff of
LNCMI. We are grateful to A. Rizzo for discussions and useful
suggestions on the manuscript, and T. Achilli for contributing to
the Faraday measurements as a summer student. We acknowledge the
support of the \textit{Fondation pour la recherche IXCORE} and the
ANR-Programme non Th\'ematique (Grant No. ANR-BLAN06-3-139634).

\end{acknowledgments}

\bibliography{apssamp}

\end{document}